\documentclass[11pt,a4paper]{article}

\usepackage{amsmath,amssymb,amsthm,graphicx,braket,multirow,authblk,amsthm,dcolumn,latexsym,subcaption,mathtools,cite}
\usepackage[a4paper]{geometry}
\usepackage{dcolumn,latexsym}
\usepackage[normalem]{ulem}
\usepackage[utf8]{inputenc}
\usepackage[T1]{fontenc}
\usepackage{hyperref}
\usepackage{multirow}
\usepackage{array}
\usepackage{tabularx}
\usepackage{makecell}
\usepackage[table]{xcolor}
\usepackage{graphicx}
\usepackage{bbm}

\usepackage{float}

\DeclareSymbolFontAlphabet{\amsmathbb}{AMSb}%
\def\<{\langle}
\def\>{\rangle}
\newtheorem{Theorem}{Theorem}[section]
\newtheorem{Proposition}{Proposition}[section]

\newtheorem{Remark}{Remark}
\theoremstyle{definition}
\newtheorem{Example}{Example}
\newtheorem{CON}{Conjecture}
\newtheorem{Cor}{Corollary}

\def\oper{{\mathchoice{\rm 1\mskip-4mu l}{\rm 1\mskip-4mu l}
		{\rm 1\mskip-4.5mu l}{\rm 1\mskip-5mu l}}}

\newcommand{\e}{\mathrm{e}}

\usepackage{hyperref}
\hypersetup{colorlinks}
\usepackage{xcolor}
\definecolor{cblue}{rgb}{0.16, 0.32, 0.75}
\definecolor{cred}{rgb}{0.7, 0.11, 0.11}
\hypersetup{%
	,linkcolor=cred
	,citecolor=cblue
	,urlcolor=black
}

\begin{document}

\title{ \bf Spectral Delineation of Markov Generators: \\  Classical vs Quantum
}
	
	

	
\author {Dariusz Chru\'sci\'nski$^1$, Sergey Denisov$^{2}$, Wojciech Tarnowski$^4$, and Karol \.Zyczkowski$^{4,5}$ \\
$^1$ Institute of Physics, Faculty of Physics, Astronomy and Informatics, Nicolaus Copernicus University,  
87-100 Toru\'n, Poland \\
	$^2$Department of Computer Science, Oslo Metropolitan University, N–0130 Oslo, Norway \\
$^4$Institute of Theoretical Physics, Jagiellonian University,
30–348 Cracow, Poland\\
$^5$Center for Theoretical Physics, 
Polish Academy of Sciences,
02–668 Warsaw, Poland}

	\maketitle


	
\begin{abstract}
The celebrated theorem of Perron and Frobenius implies
that spectra of classical Markov operators,
represented by stochastic matrices, are restricted to
the unit disk. This property holds also
for spectra of quantum stochastic maps (quantum channels), which describe quantum Markovian evolution in discrete time. Moreover, the spectra of stochastic $N \times N$ matrices are additionally restricted to a subset of the
unit disk, called Karpelevi\u{c} region, the shape of which depends on $N$.  We address the question of whether the spectra of generators, which induce Markovian evolution in continuous time, can be bound in a similar way. We propose a rescaling that allows us to answer this question affirmatively. The eigenvalues of the rescaled classical generators are confined to the modified Karpelevi\u{c} regions, whereas the eigenvalues of the rescaled quantum generators fill the entire unit disk. 
\end{abstract}
	

\section{Introduction}

Markov processes play an important role in various branches of modern science such as  physics, chemistry, biology, economy, and finance~\cite{Kampen,MM2}. 
This is due to the remarkable modeling power of Markov processes, which provide a flexible mathematical framework for describing natural systems that evolve probabilistically over time. Conventionally, these processes are divided into two families, time-discrete and time-continuous, depending on whether the  evolution occurs at discrete steps or continuously over time.

In the classical limit, discrete-time Markovian evolution is governed by stochastic matrices~\cite{Metzler-1,P2}, while in the quantum setting it is described by quantum stochastic maps,  
also known as "completely positive trace-preserving (CPTP) maps"~\cite{Paulsen,Stormer,Bhatia,ALICKI} and "quantum channels"~\cite{watrous2018,holevo}. Spectra of both families of operators carry essential characteristics that reflect im
portant features of induced evolution such as the speed of relaxation to the asymptotic state(s)~\cite{LPW2009}, the existence of quasi-stationary states~\cite{Collet_2013}, and phenomena like the cutoff effect~\cite{Diaconis,Kastor}.

According to the celebrated Perron-Frobenius theorem~\cite{Minc,Metzler-2,Metzler-1,P1,P2}, the spectrum $\sigma(S) = \{\lambda_0,\lambda_1,\ldots,\lambda_{N-1}\} \subset \mathbb{C}$ of any $N \times N$ stochastic matrix $S$ has the following properties: The spectral radius $\rho(S) = \lambda_0 = 1$, that is,  all remaining eigenvalues belong to the unit disk, i.e. $|\lambda_k|\leq 1$. Moreover, the spectrum $\sigma(S)$ is symmetric w.r.t. the real line, that is, it is invariant invariant under the complex conjugation. The eigenvector corresponding to $\lambda_0$ defines (up to normalization) a probability vector being an invariant state of $S$.

Interestingly, for a given $N$, not all points on the unit disk can be eigenvalues of $N \times N$ stochastic matrices. Indeed, if $N=2$, the spectrum is real, i.e., $\lambda_0 = 1$ and $\lambda_1 \in [-1,1]$. It was Kolmogorov (see introduction in Ref.~\cite{DD}) who first asked under what conditions a given complex number $z$ can be an eigenvalue of a some $N \times N$  stochastic matrix $S$. The question was answered by Dmitriev and Dynkin for $N \leqslant 5$~\cite{DD}, and later the proof was generalized by Karpelevi\u{c} for arbitrary $N$~\cite{Karpelevich}.

\begin{figure}[t]
\begin{center}
\includegraphics[width=0.99\textwidth]{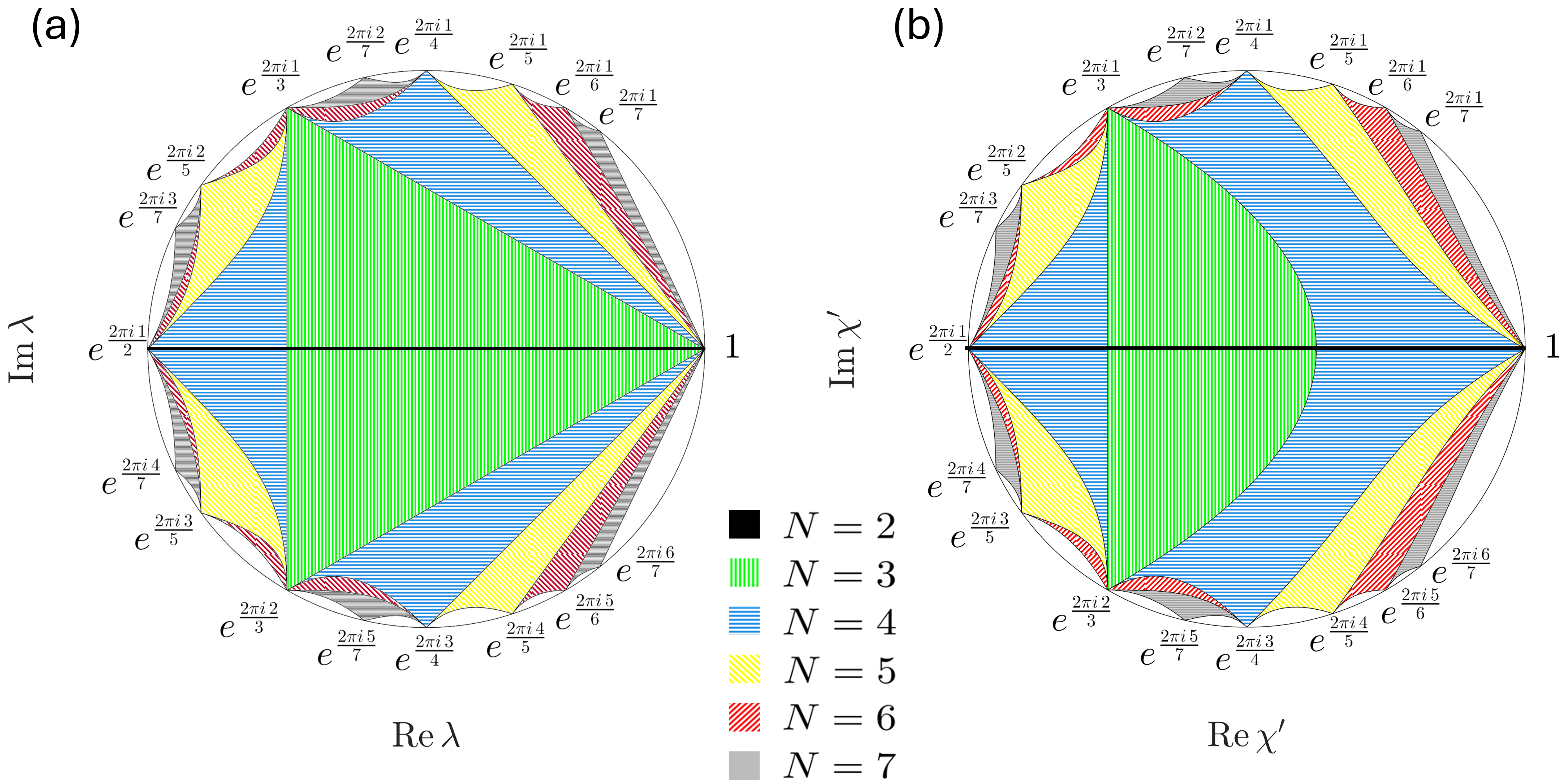}
\caption{ (a) Karpelevi\u{c} regions $\Theta_N$  bound the spectra of stochastic matrices of size $N$. Here we show the regions for $N$ ranging from $N=2$ to $N=7$ .  
(b) Modified Karpelevi\u{c} regions $\tilde{\Theta}_N$ bound the spectra of rescaled (and shifted by $\oper_N$) Kolmogorov operators. For each dimension $N$, $\tilde{\Theta}_N$  forms a subset of $\Theta_N$. In the limit $N \to \infty$, both $\Theta_N$  and $\tilde{\Theta}_N$ fill the entire unit disk.
\label{F-0}}
\end{center}
\end{figure}

According to Dmitriev, Dynkin, and Karpelevi\u{c}, the  set
of points on complex plane 
\begin{equation}\label{}
  \Theta_N = \{ z \in \mathbb{C} \mid z ~ \text{is an eigenvalue of an}~ N \times N ~ \text{stochastic matrix} \},
\end{equation}
forms a geometric region that is a proper closed subset of the unit disk, with $\Theta_N \subseteq \Theta_{N+1}$; see Figure~\ref{F-0}(a). 
These \textit{Karpelevi\u{c} regions}
are reviewed in more detail in the next section.

In the quantum case, the situation is different. The quantum analog of the Perron-Frobenius theorem \cite{Perron-1,Perron-2} implies that the spectrum of any completely positive trace-preserving map (CPTP) ~\cite{Paulsen,Stormer,Bhatia} $\sigma(\Phi) = \{\mu_0,\mu_1,\ldots,\mu_{N^2-1}\}$ is confined in the unit disk $\mathbb{D}(0,1)$ with the Perron-Frobenius eigenvalue $\mu_0=1$. In contrast to the classical case, any point on the disk can serve as an eigenvalue of some CPTP map $\Phi : {\mathcal M}_N(\mathbb{C}) \to {\mathcal M}_N(\mathbb{C})$ (we present a constructive proof in the beginning of Section~\ref{Lindblad}). In fact, the same result applies to positive trace-preserving maps~\cite{WOLF}.

The restriction of the eigenvalues of Markov generators to the unit disk $\mathbb{D}(0,1)$ can be interpreted as a \textit{delineation}, meaning that the eigenvalues can be restricted to the geometrical set on the complex plane with a well-defined  boundary. In addition, the eigenvalues of the stochastic matrices $N \times N$ are restricted to the  Karpelevi\u{c} regions~\cite{Karpelevich}; see Figure~1(a). In this work, we address the question of whether a similar delineation can be performed for the generators of time-continuous Markovian evolutions:
the classical operator ${\cal K}$ 
of Kolmogorov and the
quantum operator $\mathcal{L}$ of Lindblad.

Consider a classical system with finite number $N$ of states. Let $p_k(t)$ denote the probability to find the system in state $k$. Then the probability vector $\mathbf{p} =(p_1,\ldots,p_N)^T$ evolves according to the well known classical master  equation,
\begin{equation}\label{CM}
  \dot{p}_i(t) = \sum_{j=1}^N {\cal K}_{ij} p_j(t) , ~~i=1,\ldots, N ,
\end{equation}
where the $N \times N$ real matrix ${\cal K}$ (henceforth we will also address it as 'Kolmogorov operator') satisfies the following conditions \cite{Kampen}:
\begin{equation}\label{KK}
  {\cal K}_{ij} \geq 0  \ \ ({\rm for } \ i\neq j) \ , \ \ \ \sum_{i=1}^N {\cal K}_{ij}= 0 . 
\end{equation}
Any such operator can be represented as
\begin{equation}\label{KW}
  {\cal K}_{ij} = W_{ij} - \delta_{ij} w_j \ , \ \ \ w_j = \sum_k W_{kj} ,
\end{equation}
where $W$ is an $N \times N$ matrix with non-negative off-diagonal elements ($W_{ij}$ defines a transition rate $j \to i$). Note that diagonal elements $W_{ii}$ do not influence ${\cal K}_{ij}$. A matrix $W$ is often referred to as a \textit{Metzler matrix}~\cite{Metzler-1,Metzler-2,Minc}.

Using Eq.~(\ref{KW}), the original equation (\ref{CM}) can be recast in the form of the classical Pauli rate equation~\cite{Kampen},
\begin{equation}\label{}
  \dot{p}_i(t) = \sum_{j=1}^N \Big( W_{ij} p_j(t) - W_{ji} p_i(t) \Big).
\end{equation}
The normalization condition  $\sum_i  {\cal K}_{ij}= 0$  guarantees that $\sum_i p_i(t)=1$ for all $t \geq 0$. The corresponding solution reads $\mathbf{p}(t) = T(t)\mathbf{p}_0 $, where $T(t) = e^{{\cal K} t}$ defines a family of columnwise stochastic matrices, $T_{ij}(t) \geq 0$ and $\sum_i T_{ij}(t) = 1$. The family $\{T(t)\}_{t \geq 0}$ (classical dynamical map) satisfies a semigroup composition law $T(t+s) = T(t)T(s) = T(s)T(t)$ and $T(0) = \oper_N$, where  $\oper_N$ is $N \times N$ identity matrix.

Quantum dynamical semigroup \cite{ALICKI} is represented by a dynamical map $\{\Lambda_t\}_{t \geq 0}$ satisfying a semigroup composition law $\Lambda_t \Lambda_s = \Lambda_{t+s}$. Each map $\Lambda_t : {\mathcal M}_N(\mathbb{C}) \to {\mathcal M}_N(\mathbb{C})$ is completely positive and trace-preserving along with the initial condition, $\Lambda_0 = {\rm id}_N$,
where ${\rm id}_N$ denotes the identity map in the matrix algebra ${\mathcal M}_N(\mathbb{C})$. The quantum counterpart of Eq.~(\ref{CM}) describes Markovian evolution of the density operator $\rho_t$, i.e. it is governed by the quatum Master Equation  $\dot{\rho}_t = \mathcal{L}(\rho_t)$,  where the corresponding generator $\mathcal{L}:~{\mathcal M}_N(\mathbb{C}) \to {\mathcal M}_N(\mathbb{C})$ is represented in the celebrated Gorini-Kossakowski-Lindblad-Sudarshan (GKLS) form \cite{GKS,L} (throughout this work, we use natural units where $\hbar=1$):
\begin{equation}\label{GKLS}
  \mathcal{L}(\rho) = -i[H,\rho] + \sum_k \gamma_k \Big( L_k \rho L_k^\dagger - \frac 12 \{ L_k^\dagger L_k, \rho\} \Big),
\end{equation}
where $H$ is an Hermitian operator (effective system's Hamiltonian), $L_k$ are the so-called jump operators, and $\gamma_k > 0$ are the damping rates. Henceforth we will also refer to $\mathcal{L}$ as the `Lindblad operator'. 
The evolution of the density matrix is given by $\rho_t = \Lambda_t(\rho_0)$, where $\Lambda_t = e^{t\mathcal{L}}$.  

It is well known that the spectrum $\sigma({\cal K}) = \{\chi_0=0,\chi_1,\ldots,\chi_{N-1}\} \subset \mathbb{C}$ is symmetric w.r.t. real line, i.e.  invariant under the complex conjugation,  and ${\rm Re}\,\chi_n \leq 0$. It is clear that, since ${\cal K}$ can be multiplied by any positive real number, the spectra of $N \times N$ Kolmogorov operators can cover the entire semi-plane $\text{Re}\, \chi \leqslant 0$. In this paper we introduce a rescaling procedure which guarantees that the spectrum of the rescaled operator 
\begin{equation}
    \mathcal{K} \to \widetilde{ \mathcal{K}} = \frac{1}{\nu_{\rm c}} \,  \mathcal{K} , 
\end{equation}   
where $\nu_{\rm c}$ denotes a {\em classical} rescaling parameter, i.e. 
$\sigma(\tilde{{\cal K}}) =\{\tilde{\chi}_0=0,\tilde{\chi}_1,\ldots,\tilde{\chi}_{N-1}\} \subset \mathbb{C}$, is constrained to the unit disk $\mathbb{D}(-1,1)$
centered at $\chi=-1$. Moreover, for a
given dimension $N$, the spectrum of the rescaled operator additionally shifted by one, $\sigma(\tilde{{\cal K}}) + 1 = \{\chi'_0=1,\chi'_1,\ldots,\chi'_{N-1}\}$, is confined to the geometric set $\widetilde{\Theta}_N$ which is a modification of the Karpelevi\u{c} region $\Theta_N$; see Figure~\ref{F-0}(b).

The spectrum of a GKLS generator $\mathcal{L}$, $\sigma({\cal L}) = \{\ell_0=0,\ell_1,\ldots,\ell_{N^2-1}\} \subset \mathbb{C}$, is also symmetric w.r.t. the real line and located on the non-positive semi-plane (the real parts of the eigenvalues are never strictly positive). 
In this work we show that, after appropriate rescaling
\begin{equation}
    \mathcal{L} \to \widetilde{ \mathcal{L}} = \frac{1}{\nu_{\rm q}} \,  \mathcal{L} , 
\end{equation}   
where $\nu_{\rm q}$ denotes a {\em quantum } rescaling parameter, 
the spectrum of a purely dissipative, i.e., with $H = 0$, Lindblad operator, Eq.~(\ref{GKLS}), is confined to the unit disk $\mathbb{D}(-1,1)$. 
In what follows we propose (Conjectures \ref{CON-0}--\ref{CON-3}) three different rescaling parameters $\{\nu_{\text{q},k}\}$ (with $k=1,2,3$) and conjecture that
$\nu_{\text{q},3} \leq \nu_{\text{q},2} \leq \nu_{\text{q},1}$. This means that the rescaling with $\nu_{\text{q},3}$ provides the tightest delineation.  We demonstrate that the conjectures are satisfied for important classes of GKLS generators well studied in the literature. In particular, they hold for a family of the so-called Davies generators, which are quantum Markov generators derived in the weak coupling limit~\cite{Davies-1,ALICKI,Breuer,rivas,Bassano}.

The remaining parts of the paper are organized as follows: In the next section, we review  Karpelevi\u{c}’s celebrated results characterizing the set $\Theta_N$ of admissible eigenvalues of $N \times N$ stochastic matrices. In Section~\ref{Generator} we introduce a rescaling for Kolmogorov operators and analyze the corresponding delineation  of their spectra. The quantum version of this problem is discussed in Section~\ref{Lindblad}. In Section~\ref{Beyond} we discuss a generalization to the case of quantum generators of  positive semigroups.
Section \ref{Sec-NONM} shows that for time-dependent generators giving rise to non-Markovian evolution one cannot in general bound their spectra within $\mathbb{D}(-1,1)$. Hence, the existence of the scaling is a characteristic property of Markovian dynamics. A notion of a `random  Markov generator'~\cite{timm,Random-1,Random-2,Ca19,COOG19,SRP20} and the spectral densities of the corresponding operators after the rescaling are discussed in Section~\ref{Sec_typical}. The final conclusions and outlook are presented in Section~\ref{Conclusions}.
A more detailed discussion of the maps
acting on a $N$-dimensional system is provided
in an Appendix.

\section{Admissible spectra of stochastic matrices: Karpelevi\u{c} regions
}   \label{Karpelevich}

The question posed by Kolmogorov in 1938, 
\textit{``What is the set of all complex numbers that are eigenvalues of $N \times N$ stochastic matrices?''}, was first addressed by Dmitriev and Dynkin~\cite{DD}. They reformulated the question in an elegant geometric way
and answered it for $N \leqslant 5$. They further conjectured that their results extend to arbitrary $N$. It was later proved by Karpelevi\u{c}~\cite{Karpelevich}.

The original proof is lengthy and intricate; see also Ref.~\cite{Minc}. Here, we present Karpelevi\u{c}'s results in the form proposed by Ito~\cite{Ito}, while also referring to recent works~\cite{JP,Smigoc}.
Following these works,  we address row stochastic matrices.

Given integer $N$, one calls  the set $F_N := \{ p/q\, |\, 0\leq p < q \leq N\}$ with $p$ and $q$ being co-prime, the series of Farey fractions of order $N$~\cite{Hardy}. We start by presenting the following theorem~\cite{Ito}:

\begin{Theorem} The region $\Theta_N$ intersects the unit circle $\{ z \in \mathbb{C}\,|\,, |z|=1\}$ at the points $\{ e^{2\pi i p/q}\, |\, p/q \in F_N\}$. The boundary of geometric set $\Theta_N$ consists of points $e^{2\pi i p/q}$ and curvilinear arcs connecting them in circular order. Let the endpoints of an arc be   $e^{2\pi i p/q}$ and $e^{2\pi i r/s}$ with $q<s$. Each of these arcs is given by the following parametric equation

\begin{equation}\label{K!}
  z^s (z^q - \beta)^{\lfloor N/q \rfloor} = \alpha^{\lfloor N/q \rfloor} z^{q \lfloor N/q \rfloor} ,
\end{equation}
where $z \in \mathbb{C}$, the real parameter $\alpha \in [0,1]$ and $\beta =1-\alpha$ ($ \lfloor x \rfloor$ denotes the floor function; for $x \geq 0$ it is just an integer part of $x$).
\end{Theorem}

Recently, Johnson and Paparella~\cite{JP} discussed  families of stochastic matrices that realize each of the arcs. Later, it was observed~\cite{Smigoc} that these families are not unique, and further insights into the structure of realizing stochastic matrices were obtained using a graph-based approach. It was proved~\cite{Karpelevich,JP,Smigoc} that there are four families of arcs:

\begin{itemize}
  \item Type 0: $q=1$, $s=N$
  \item Type I: $\lfloor N/q \rfloor = 1$, $s=N$
  \item Type II: $\lfloor N/q \rfloor > 1$ and $s < q \lfloor N/q \rfloor$,
  \item Type III: $\lfloor N/q \rfloor > 1$ and $s=N >  q \lfloor N/q \rfloor$.
\end{itemize} 
Denote by $C_N$ a circulant matrix,

\begin{equation}\label{}
C_N = \begin{pmatrix} 
                0 & \oper_{N-1} \\
                1 & \boldsymbol{0} 
              \end{pmatrix} .
\end{equation}
Stochastic  matrices $M(\alpha)$ of order $N$ realizing the above arcs can be constructed as follows \cite{JP,Smigoc}:

\begin{itemize}
  \item {\it Type 0 arc}.  Equation (\ref{K!}) reduces to

\begin{equation}\label{K0}
  (z-\beta)^N - \alpha^N = 0.
\end{equation}
The corresponding family of stochastic matrices realizing 
Eq. (\ref{K0}) reads \cite{JP},

\begin{equation}\label{M-0}
  M(\alpha) = \alpha C_N + \beta \oper_N  . 
\end{equation}

  \item {\it Type I arc}:  $\lfloor N/q \rfloor =1$. Equation (\ref{K!}) reduces to

\begin{equation}\label{KI}
  z^N - \beta z^{N-q} - \alpha = 0 ,
\end{equation}
and the corresponding family of $N \times N$ stochastic matrices realizing (\ref{KI}) has the following form \cite{JP}

\begin{equation}\label{}
  M(\alpha) = \begin{pmatrix} 
                0 & \oper_{N-1} \\
                \alpha & \boldsymbol{\beta} 
              \end{pmatrix} ,
\end{equation}
where the vector $\boldsymbol{\beta} \in \mathbb{R}^{N-1}$ has only one non-zero component $(\boldsymbol{\beta})_{N-q}=\beta$. 

  \item {\it Type II arc}: $\lfloor N/q \rfloor > 1$ and $s < q \lfloor N/q \rfloor$. Equation (\ref{K!}) reduces to

\begin{equation}\label{KII}
  (z^q -\beta)^{\lfloor N/q \rfloor} - \alpha^{\lfloor N/q \rfloor} z^{q\lfloor N/q \rfloor - s}  = 0 .
\end{equation}
The corresponding family $M(\alpha) = \alpha X + \beta Y$, where $X$ is the nonegative companion matrix of the polynomial $z^{q\lfloor N/q \rfloor} - z^{q\lfloor N/q \rfloor -s}$, and

\begin{equation}\label{}
  Y = \bigoplus_{k=1}^{\lfloor N/q \rfloor} C_q , 
\end{equation}
is a $q\lfloor N/q \rfloor \times q\lfloor N/q \rfloor$ stochastic matrix. Recall that a companion matrix of the polynomial $c_0 + c_1 t + \ldots c_{N-1} z^{N-1} + z^N$ is defined by

\begin{equation}\label{}
  X = \begin{pmatrix} 
                0 & \oper_{N-1} \\
                -c_0 & \boldsymbol{c} 
              \end{pmatrix} ,
\end{equation}
with ${\bf c} =(-c_1,\ldots,-c_{N-1})$.


  \item {\it Type III arc}: $\lfloor N/q \rfloor > 1$ and $s=N > q \lfloor N/q \rfloor$. Equation (\ref{K!}) reduces to

\begin{equation}\label{KIII}
  z^d(z^q -\beta)^{\lfloor N/q \rfloor} - \alpha^{\lfloor N/q \rfloor}  = 0 ,
\end{equation}
where $d := N - q \lfloor N/q \rfloor > 0$. The realizing family has the following form \cite{JP} 

\begin{equation}
 M(\alpha) = \alpha C_N + \beta Y   ,
\end{equation}
where

\begin{equation}\label{}
  Y = J_d(0) \oplus C_q \oplus \ldots \oplus C_q , 
\end{equation}
and $J_d(\lambda)$ denotes $d\times d$ Jordan block.


\end{itemize}

\begin{Example} For $N=3$ one has the Farey sequence~\cite{Hardy}
of length four:

$$ F_3 =\left\{ 0, \frac 13, \frac 12, \frac 23 \right\} . $$
One has the following arcs parameterized by $\alpha \in [0,1]$ (recall, that  $\beta = 1-\alpha$):
\begin{itemize}
  \item Type 0 arc: It connects $z=1$ and $z=e^{\pm 2\pi i /3}$. From Eq.~(\ref{K0}) one gets $(z-\beta)^3 = \alpha^3$ which implies apart from $z=1$ two straight lines 

\begin{equation}\label{}
   \ z(\alpha) = \beta + \alpha e^{2\pi i /3} \ , \ z(\alpha) = \beta +  \alpha e^{- 2\pi i /3} .
\end{equation}

\item Type I arc: $\lfloor 3/q \rfloor =1$ and $q < s=3$ implies $q=2$.  From Eq.~(\ref{KI}) one obtains $z^3-\beta z - \alpha=0$. One finds

$$  z^3-\beta z - \alpha = (z-1)(z^2+z+\alpha) , $$
and hence apart form $z=1$ one obtains

\begin{equation}\label{}
  z(\alpha) = \frac{-1 \pm \sqrt{1 - 4 \alpha}}{2} .
\end{equation}
For $\alpha \leq 1/4$ one has $z(\alpha) \in [-1,0]$ and for $\alpha > 1/4$

\begin{equation}\label{}
  z(\alpha) = - \frac 12 \pm i\, \frac{\sqrt{4\alpha - 1}}{2} ,
\end{equation}
defines a vertical line connecting $e^{2\pi i/3}$ and $e^{-2\pi i /3}$ corresponding to $\alpha =1$. The structure of $\Theta_3$, that is a triangle with a 'tail' formed by $\Theta_2$ [see  Figure \ref{F-0}(a)],  was first found by Dmitriev and Dynkin~\cite{DD}.

\end{itemize}
For $N=3$ there are neither Type II  nor Type III  arcs.
\end{Example}

\begin{Example}
 For $N=4$ we have the following Farey sequence:
$$ F_4 =\left\{ 0, \frac 14, \frac 13, \frac 12, \frac 23,\frac 34 \right\} . $$
One has the following arcs parameterized by $\alpha \in [0,1]$ (recall, that  $\beta = 1-\alpha$):
\begin{itemize}
  \item Type 0 arc connects $z=1$ and $z=e^{\pm \pi i /2}$. From Eq.~(\ref{K0}) one gets $(z-\beta)^4 = \alpha^4$ which implies apart from $z=1$ and the segment $[-1,1] \subset \mathbb{R}$, two lines 

\begin{equation}\label{}
   \ z(\alpha) = \beta + \alpha e^{\pi i /2} \ , \ z(\alpha) = \beta +  \alpha e^{- \pi i /2} .
\end{equation}

\item Type I arc: $\lfloor 4/q\rfloor  = 1$ implies $q=3$ and $s=4$. 
From  Eq. (\ref{KI}) one obtains,

\begin{equation}\label{}
  z^4 - \beta z - \alpha = (z-1)(z^3+z^2+z + \alpha)=0 .
\end{equation}
Again, apart form $z=1$ there are three solutions parameterized by $\alpha$: a segment $[-1,0] \subset \mathbb{R}$ and two lines connecting $(e^{i \pi/2},e^{i 2\pi/3})$ and $(e^{-i \pi/2},e^{-i 2\pi/3})$ (conjugated pair), see  Figure~\ref{F-0}(a).

\item Type II arc: $\lfloor 4/q\rfloor  > 1$ and $s < q  \lfloor 4/q\rfloor$. Hence $q=2$ and $s= 3$. 
Eq.~(\ref{KII}) implies that
\begin{equation}\label{KII}
  (z^2 -\beta)^2 - \alpha^2 z  = (z-1)\Big( (z+1)^2(z-1) + 2 \alpha (z+1) - \alpha^2\Big) = 0  .
\end{equation}
Apart form $z=1$ there are three solutions parameterized by $\alpha$: a segment $[0,1] \subset \mathbb{R}$ and two lines connecting $(-1,e^{i 2\pi/3})$ and $(-1,e^{-i 2\pi/3})$ (conjugated pair), see  Figure~\ref{F-0}(a)

\end{itemize}
For $N=4$ there are no Type III arcs (they appear starting from $N=5$).
\end{Example}

Interestingly, some Karpelevi\u{c} arcs arise as powers of other arcs \cite{JP}: if $S \subset \mathbb{C}$, then one defines its $m$-th power
$S^m = \{z^m \in \mathbb{C} \, |\, z\in S \}$ --
for a recent discussion see Ref.~\cite{Smigoc-3}.

\section{Bounding spectra of classical Markov generators: \\ modified Karpelevi\u{c} regions}   \label{Generator}


Let $ {\cal K}$ be a classical Markov generator. Since diagonal elements $ {\cal K}_{ii} \leq 0$, we define
\begin{equation}
\label{mudef}
\nu_{\text{c}} := \max_i |\mathcal{K}_{ii}| .    
\end{equation}
 We have, therefore, 

\begin{equation}\label{}
  {\cal K}_{ij} = {\cal K}_{ij} + \nu_{\text{c}} \delta_{ij} - \nu_{\text{c}}\delta_{ij} = W_{ij} - \nu_{\text{c}} \delta_{ij} ,  
\end{equation}
where

\begin{equation}\label{}
  W_{ij} := {\cal K}_{ij} + \nu_{\text{c}} \delta_{ij} ,
\end{equation}
and hence

\begin{equation}\label{}
  {\cal K}_{ij} =  \nu_{\text{c}} \Big( \nu_{\text{c}}^{-1} W_{ij} - \delta_{ij} \Big) .
\end{equation}
It is evident that 

\begin{equation}
    \sigma\left(\frac 1 r \mathcal{K}\right) \subset \mathbb{D}\left(-1,\frac{\nu_{\rm c}}{r}\right) , 
\end{equation}
and hence if $r > \nu_{\rm c}$, then the spectrum of $\frac 1r \mathcal{K}$ is confined to $\mathbb{D}(-1,1)$ as well.

Note that $W_{ij} \geq 0$ and $\sum_i W_{ij} = \nu_{\text{c}}$. Hence $T_{ij} =  \nu_{\text{c}}^{-1} W_{ij}$ defines a stochastic matrix. Thus the rescaled generator $\widetilde{{\cal K}} := \nu_{\text{c}}^{-1} {\cal K}$ has the form $\widetilde{{\cal K}}_{ij} = T_{ij} - \delta_{ij}$. It is, therefore clear that the spectrum of $\sigma(\widetilde{{\cal K}}) = \lbrace \tilde{\chi}_0, \tilde{\chi}_1,...,\tilde{\chi}_{N-1} \rbrace = \sigma(T) - 1$. In particular $\sigma(\widetilde{K})$ belongs to the unit disk  $\mathbb{D}(-1,1)$, while the spectrum of $T$, $\lbrace \chi'_0, \chi'_1,...,\chi'_{N-1} \rbrace$, $\chi'_n = \tilde{\chi}_n + 1$, is confined within $\mathbb{D}(0,1)$.

Since $T$ is a stochastic matrix, it is evident that $\sigma(T) \subseteq \Theta_N$. However, $T$ is not an arbitrary stochastic matrix. By construction, it satisfies the condition $\min_i T_{ii} = 0$. Does this affect the admissible region? Interestingly, it does.

Note that stochastic matrices $M(\alpha)$ realizing arcs I, II, and III satisfy $M_{ii}(\alpha) = 0$~~\cite{JP,Smigoc} and hence they already belong to our restricted class. However, this is not true for the family that realizes a Type 0 arc since $M_{ii}(\alpha)=\beta$, see Eq.~(\ref{M-0}). We consider, therefore, the following restricted class: instead of 

\begin{equation}
    M(\alpha) = \beta \oper_N + \alpha C_N = {\rm Diag}[\underbrace{\beta,\beta,\ldots,\beta}_{N}] + \alpha C_N , 
\end{equation}
we define

\begin{equation}\label{M-0t}
 \widetilde{M}(\alpha)  = {\rm Diag}[0,\underbrace{\beta,\ldots,\beta}_{N-1}] + \alpha C_N  ,
\end{equation}
%
which satisfies $ \widetilde{M}_{11}(\alpha)=0$. The eigenvalues of the modified matrix $\widetilde{M}(\alpha)$ satisfy 

\begin{equation}\label{K0t}
  z(z-\beta)^{n-1}- \alpha^{n-1} = 0 ,
\end{equation}
which replaces Eq.~(\ref{K0}).

\begin{Example} For $N=3$ equation (\ref{K0t}) reduces to $ z(z-\beta)^{2}- \alpha^{2} = 0 $ which is equivalent to
  
\begin{equation}\label{K0t}
 (z-1)\Big(z^2 + (2\alpha -1) + \alpha^2 \Big) = 0 .
\end{equation}  
Apart form $z=1$, its solution is

\begin{equation}\label{}
  z(\alpha) = \frac{1-2\alpha \pm \sqrt{1-4\alpha}}{2}  , \ \ \ \alpha \in [0,1] .
\end{equation}
For $\alpha \in [0,1/4]$ it defines a segments $[0,1] \subset \mathbb{R}$, and for $\alpha \in [1/4,1]$ it defines Type $\widetilde{0}$ arc 

\begin{equation}\label{}
  z(\alpha) = \frac{1-2\alpha \pm i\sqrt{4\alpha-1}}{2}  , 
\end{equation}
connecting $e^{\pm 2\pi i/3}$. Therefore, the spectrum of the rescaled 3-dimensional classical Markovian generator lives in the modified shape $\widetilde{\Theta}_3$ shifted by $-1$; see Figure~\ref{F-0}(b).
\end{Example}

\begin{Example} For $N=4$ equation (\ref{K0t}) reduces to 

\begin{equation}\label{}
 z(z-\beta)^{3}- \alpha^{3} = 0  = 0 .
\end{equation}
Now, apart form $z=1$, it gives rise to three families of solutions  $z(\alpha)$. Since they are quite involved (being solutions of $3$-rd order equation), we skip the analytical form. One of them is purely real and connects $z=0$ with $z=-1$. The complex conjugate pair of curves connects $z=1$ with $z=\pm i$ and represent  modified Type $\widetilde{0}$ arcs. 
Therefore, the entire spectrum of the rescaled $4$-dimensional classical Markovian generator is confined to the modified Karpelevi\u{c} set $\widetilde{\Theta}_4$, shifted by $-1$; see Figure~\ref{F-0}(b).
\end{Example}

Note that the scaling parameter $\nu_{\text{c}}$ is an intrinsic property of the generator ${\cal K}$ and does not depend on the particular representation, Eq.~(\ref{KW}). Recall that for any $N \times N$ matrix $A$, the 1-norm and $\infty$-norm are defined as follows:
\begin{equation}
    \| A \|_1 := \max_j \sum_{i=1}^N |A_{ij}|  \ , \ \ \ \| A \|_\infty := \max_i \sum_{j=1}^N |A_{ij}| 
\end{equation}
Note, that $\|A\|_1 = \|A^T\|_\infty$. One has

\begin{equation}
  \|{\cal K}\|_1 = \max_j \sum_{i=1}^N |{\cal K}_{ij}| = \max_j \left( \sum_{i\neq j}^N {\cal K}_{ij} - {\cal K}_{jj} \right) = 2 \nu_{\text{c}} ,
\end{equation}
due to $ \sum_{i\neq j}^N {\cal K}_{ij}= - {\cal K}_{jj}$.

Suppose now that a specific representation is given, i.e., ${\cal K}_{ij} = W_{ij} - \delta_{ij} w_j$, where $w_j = \sum_i W_{ij}$. Now, since diagonal elements $W_{ii}$ does not affect $\mathcal{K}_{ij}$ one can always use a special {\em gauge} $W^{(0)}_{ij}$ such that $W^{(0)}_{ii} = 0$.

\begin{Proposition} \label{PRO-1} The parameters $\nu_{\text{c}}$ reads

\begin{equation}
    \nu_{\text{c}} =  \|W^{(0)}\|_1 ,
\end{equation}
i.e., $\nu_{\text{c}}$ is uniquely defined by the rate matrix $W^{(0)}_{ij}$.     
\end{Proposition}
Indeed, let us  define 

\begin{equation}\label{tilde-W}
  \widetilde{W}_{ij} := W^{(0)}_{ij} + \delta_{ij} (\|W^{(0)}\|_1 -w^{(0)}_i) .
\end{equation}
Clearly $\widetilde{W}_{ij} \geq 0$ and $\min_i \widetilde{W}_{ii}=0$. One has, therefore,

\begin{equation}\label{}
  {\cal K}_{ij} = \widetilde{W}_{ij} -  \|W^{(0)}\|_1 \delta_{ij} = \|W^{(0)}\|_1 \Big( \|W^{(0)}\|_1^{-1} \widetilde{W}_{ij} - \delta_{ij} \Big) ,
\end{equation}
which implies that $\nu_{\text{c}} =  \|W^{(0)}\|_1 =\| W^{(0)T}\|_\infty$. 

Note, that $\nu_{\text{c}}$ is a Perron-Frobenius eigenvalue of $\widetilde{W}_{ij}$ and hence defines a spectral radius of $\widetilde{W}_{ij}$. One has therefore

\begin{equation}\label{!!!}
 \nu_{\text{c}} = \lambda_{\rm PF}(\widetilde{W}) \equiv \rho(\widetilde{W}) = \|W^{(0)}\|_1  ,
\end{equation}
where $\rho(A)$ denotes a spectral radius of $A$. Equivalently one has in the dual (Heisenberg) picture

\begin{equation}\label{Wx0}
  \widetilde{W}^T \mathbf{x}_0 = \|W^{(0)T}\|_\infty \mathbf{x}_0 ,
\end{equation}
where $\mathbf{x}_0=(1,\ldots,1)^T$.

\begin{Remark} Recently, a slightly more general problem of spectral properties of Metzler matrices was analyzed in Refs.~\cite{M1,M2,M3}. Recall, that a square matrix $M$  of order $N$ is called Metzler matrix iff all its off-diagonal  elements are non-negative. It is evident that ${\cal K}$ is a particular Metzler matrix satisfying an additional constraint $\sum_i {\cal K}_{ij}=0$ for all $j=1,\ldots,N$.   
A relation between the location of the spectra of $M$ and 
the Karpelevi\u{c} regions was discussed in Refs.~\cite{M1,M2,M3}. 
\end{Remark}

\begin{Remark} The structure of spectra of Kolmogorov operators implies a  bound for a ratio  between the imaginary and real parts of the eigenvalues. A simple trigonometric analysis leads to the bound
\begin{equation}
    \Big| \frac{{\rm Im}\, \chi}{{\rm Re}\, \chi} \Big| \leq \cot{\frac{\pi}{N}} ,
\end{equation}
for any complex eigenvalue $\chi$ of a $\mathcal{K}$-operator.  This bound was recently addressed in Ref.~\cite{blum2024}, where it was found to play an important role in the analysis of population oscillations.
\end{Remark}

\begin{figure}[t]
\begin{center}
\includegraphics[width=0.99\textwidth]{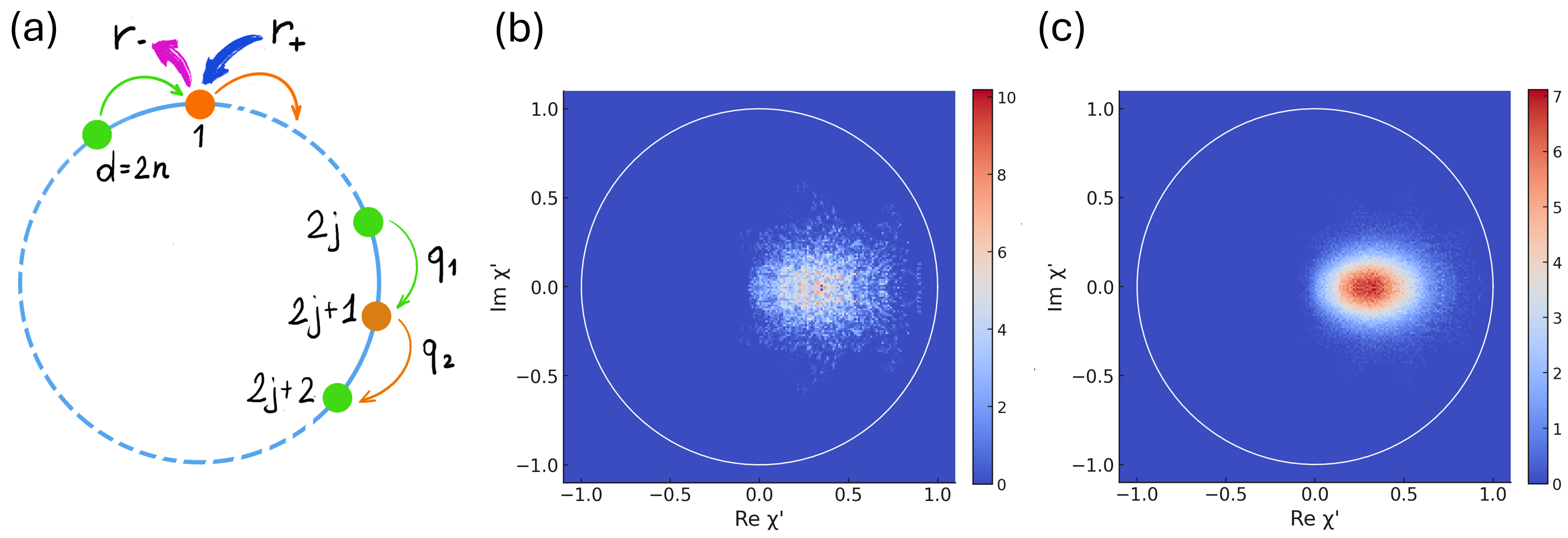}
\caption{(a) A totally asymmetric exclusion process and (b-c) the spectral densities of the rescaled Kolmogorov operators ${\cal K'_{\text{TASEP}}}$, Eq.~(\ref{eq:TASEP}), for different numbers of sites $d$ (the total number of states is $N = 2^d$). The spectral densities were obtained by randomly and independently sampling the transition rates $q_1$ and $q_2$ from the interval $[0.5, 5]$. The number of samples is $10^4$ in (b) and $10^3$ in (c). \label{F-2}}
\end{center}
\end{figure}

\subsection{Illustration:  A totally asymmetric exclusion process 
}   \label{TASEP}

The rescaling procedure introduced in the previous section allows us to compare spectra of different Kolmogorov operators, regardless of their  dimensions or/and  origins. Furthermore, this procedure enables us to compute the spectral density of the rescaled eigenvalues, $\{ \chi' \}$, for the same model by sampling across a wide range of parameter values.

To illustrate this idea, we use the continuous-time version of the totally asymmetric simple exclusion process (TASEP)~\cite{Derrida}, which we briefly outline below; see  also Figure~\ref{F-2}(a).

Consider a ring with $d = 2n$ sites,  $n = 1,2,\dots$. Particles can hop between neighboring sites in the clockwise direction only, with two different hopping rates depending on whether the site a particle hops from is labeled with an even or odd $d$. Particles can be introduced into or removed from the system at site $j = 1$, with rates $r_{\text{in}}$ and $r_{\text{out}}$, respectively. The hard-core interaction prevents more than one particle from occupying the same site, thereby reducing the total number of states to $N = 2^d$.

The Kolmogorov generator ${\cal K}$ of the TASEP process is given by  
\begin{equation}
    {\cal K_{\text{TASEP}}} = \sum_{j=1}^{d} \left[ q_1 \sigma^-_{2j-1} \sigma^+_{2j} + q_2 \sigma^-_{2j} \sigma^+_{2j+1} \right] + r^+ \sigma^+_1 + r^- \sigma^-_1,
    \label{eq:TASEP}
\end{equation}  
where $\sigma^+_j$ and $\sigma^-_j$ are the Pauli raising and lowering operators at site $j$, representing the creation and annihilation of particles at site $j$. The periodic boundary condition are imposed so that from site $2d$ particles hop to site $1$. The parameters $q_1$ and $q_2$ denote the rate of hopping from odd to even sites and from even to odd sites, respectively.  In the case $q_1=q_2=q$ the model is exactly solvable using a matrix-product ansatz~\cite{Derrida}.

We consider the non-integrable case $q_1 \neq q_2$, with both rates independently and uniformly sampled from the interval $[0.5,5]$. The results of the sampling for different values of $d$ are shown in Figure~\ref{F-2}(b-c). Notably, the  spectral densities of the rescaled generators are localized within an ellipse-like region that does not shrink substantially  with the  (exponential) increase of the dimension. This is in a strong contrast to the behavior of the spectra of rescaled random Kolmogorov operators, which exhibit spectral support quickly shrinking  as the number of states increases -- see Section~\ref{Sec_typical} for a more detailed discussion.

\section{Inverse eigenvalue problem for dissipative Lindblad generators}  \label{Lindblad}

We begin this section  by briefly outlining the spectral properties of completely positive trace-preserving (CPTP) maps acting in the space ${\mathcal M}_N(\mathbb{C})$ of $N \times N$ complex matrices~\cite{watrous2018,holevo,WOLF} (a reminder: we also refer to them as `quantum stochastic maps'). Since these maps preserve Hermiticity, their eigenvalues are either real or occur in complex conjugate pairs so the spectrum is invariant under the complex conjugation. Their trace-preserving property ensures that they always have at least one eigenvalue, $\mu_0 = 1$. Finally, since quantum stochastic maps are positive, their spectral radius is exactly one~\cite{Perron-1}. As a result, all their eigenvalues are confined within the unit disk $\mathbb{D}(0,1)$.

\begin{figure}[t]
\begin{center}
\includegraphics[width=0.85\textwidth]{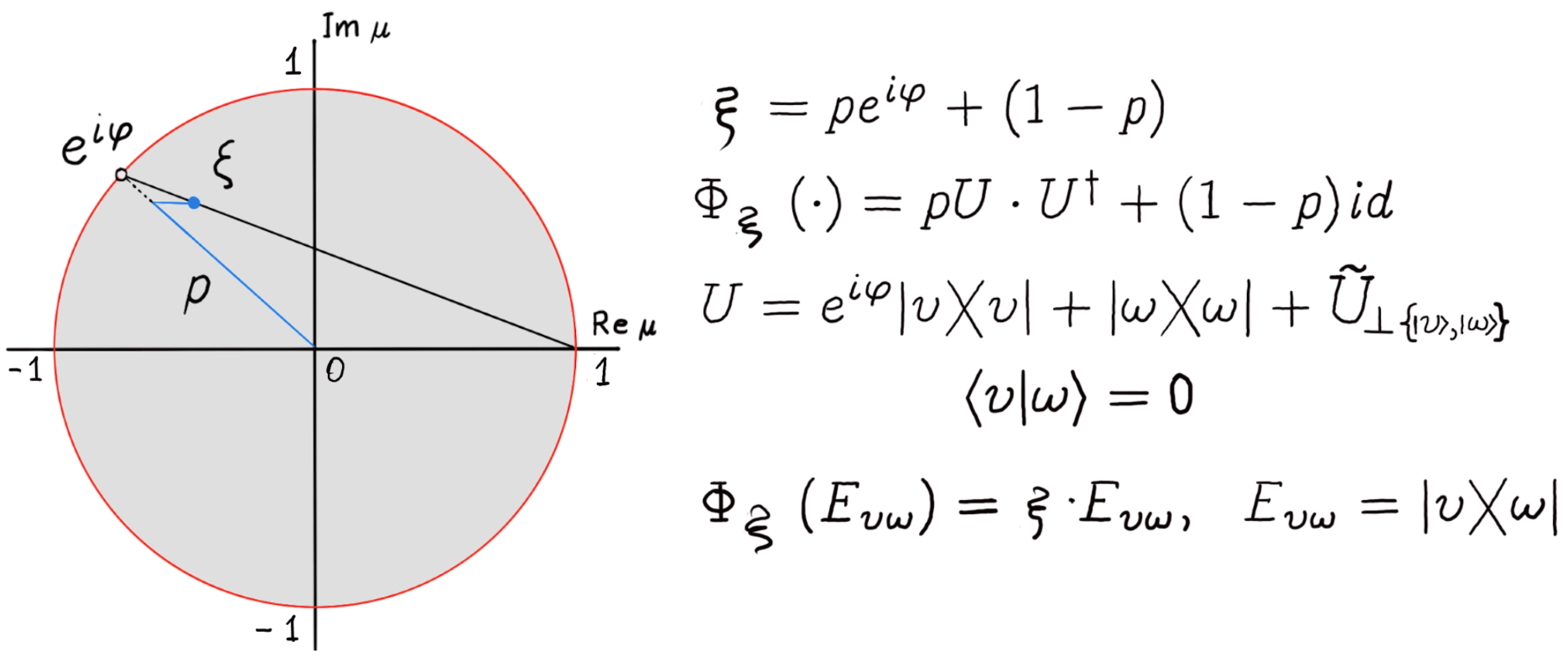}
\caption{Any point $\xi$ in the unit disk can be an eigenvalue of a CPTP map.  
Such a map, $\Phi_{\xi}$, can be explicitly constructed for any $N \geqslant 2$, by using a parametrization of points on the unit disc with $0 \leqslant p \leqslant 1$ and $\varphi \in [-\pi,\pi]$.  Here, $\tilde{U}_{\perp \{\upsilon, \omega\}}$ is a random unitary acting in the orthogonal complement  
of the two-dimensional subspace spanned by $\vert \upsilon \rangle$ and $\vert \omega \rangle$.
}  \label{F2}
\end{center}
\end{figure}

In contrast to the classical stochastic maps, there are no further restrictions on the eigenvalues of quantum stochastic maps. In other words, any point inside the unit disc can be an eigenvalue of a CPTP map. Given a complex number $\xi \in \mathbb{D}(0,1)$ and dimension $N \geqslant 2$, we can construct a map that has $\xi$ as an eigenvalue. A particular recipe 
is presented in Fig.~\ref{F2}.  The map $\Phi_{\xi}$ has $\xi$ as an eigenvalue of multiplicity one, with 
the corresponding eigenoperator $E_{\upsilon \omega}=\vert \upsilon \rangle \langle \omega\vert$. The multiplicity can be gradually increased, up to $\lfloor \frac{N-1}{2} \rfloor$, by chirping out more projectors from the unitary $\tilde{U}_{\perp \{\upsilon, \omega\}}$. The spectrum of the map also includes the conjugate complex eigenvalue, $\xi^{\ast}$, with the corresponding eigenoperator $E_{\omega \upsilon}=\vert \omega \rangle \langle \upsilon\vert$.

\begin{Remark} There is an interesting result presented in Ref.~\cite{WPG}: For any set of $d$ nonzero complex numbers, $\boldsymbol{\xi} = 
\lbrace \xi_1, \xi_2, \dots, \xi_d \rbrace$, which is confined to the unit disk,  
invariant under complex conjugation, and includes $1$, there exists a stochastic quantum map $\Phi_{\boldsymbol{\xi}}: {\mathcal M}_N(\mathbb{C}) \to {\mathcal M}_N(\mathbb{C})$, with  
$N \leqslant 2 \max \lbrace d-1,1 \rbrace$,  
such that the given set forms part of the map spectrum (while the remaining eigenvalues are zero).  In the ultimate case, all members of the set are real, with $d-1$ values different from $1$.  In this scenario, each $ \xi_j$ would require a separate map, so the dimension of the total Hilbert space,  
$\mathcal{H} := \bigoplus_j \mathcal{H}_j$, is $2(d-1)$.
\end{Remark}

Consider now  quantum Markov generator $\mathcal{L}: {\mathcal M}_N(\mathbb{C}) \to {\mathcal M}_N(\mathbb{C})$ which gives rise a to completely positive trace-preserving dynamics.
It can be represented in the well known Lindblad form, Eq.~(\ref{GKLS}). Could we expect that the spectrum of a properly rescaled $\mathcal{L}$ is also located within the unit disk shifted by $-1$?

In general, the answer is negative. Consider, for instance, a purely Hamiltonian case, i.e., $\mathcal{L}(\rho) = -i[H,\rho]$. Clearly, if $\{E_1,\ldots,E_n\}$ are the eigenvalues of $H$, then the spectrum of $\mathcal{L}$ consists of purely imaginary eigenvalues $\{\pm i (E_k - E_\ell)\}$. It is evident that this spectrum cannot be rescaled to fit within the unit disk centered at $(-1,0)$. In what follows, we focus on a purely dissipative generator, corresponding to the case $H = 0$.

Similar to the spectrum of a quantum stochastic map, the spectrum of a purely dissipative Lindblad operator $\mathcal{L}$  
is invariant under complex conjugation since $\mathcal{L}$ preserves Hermiticity.  
Because $\mathcal{L}$ generates a CPTP group, its spectrum must contain at least one eigenvalue at $0$.  Furthermore, all nonzero eigenvalues are strictly negative,  $\text{Re}\, \lambda_j < 0$.  

In contrast to the case of classical Markov generators (see Remark 2), there are no additional restrictions on the eigenvalues of the quantum Markov generators. In other words, any point in the negative half-plane can be an eigenvalue of a dissipative Lindblad operator. We discuss the explicit construction of the operator $\mathcal{L}_{\xi}$ given below, after Eq.~(\ref{E2}). 

We start by considering a simple class of Lindblad operators for which the spectrum can be restricted to the unit disk after an appropriate rescaling. Let $\Phi$ be an arbitrary  CPTP map and define
\begin{equation}   \label{E2}
    \mathcal{L}(\rho) = \gamma( \Phi(\rho) - \rho),
\end{equation}
with $\gamma >0$.  Now, since the spectrum of $\Phi$ is confined in the unit disk $\mathbb{D}(0,1)$ \cite{Perron-1,Perron-2}, the corresponding spectrum of $\frac 1 \gamma \mathcal{L}$ is located in $\mathbb{D}(-1,1)$.

Now it is evident that for any given point $\xi$ with $\text{Re} \, \xi < 0$, it is possible to construct a Lindblad operator $\mathcal{L}_{\xi}$ that has $\xi$ as its eigenvalue: The construction follows directly from the procedure for constructing a map $\Phi_{1+\xi/\gamma}$ (see Fig.~\ref{F2}) and a Lindblad operator of the form given in Eq.~(\ref{E2}). Note that one must first construct a map for an eigenvalue $1 + \frac{\xi}{\gamma}$, which requires choosing $\gamma$ such that $\gamma > \frac{\xi}{2}$.

As a special example of Eq.~(\ref{E2}), let us consider
\begin{equation}  \label{E1}
    \mathcal{L}(\rho) = \gamma( U \rho U^\dagger - \rho) ,
\end{equation}
where $U$ is an arbitrary $N \times N$ unitary matrix and $\gamma> 0$. If $e^{i\phi_k}$ are eigenvalues of $U$, then $e^{i(\phi_k - \phi_\ell)}-1$ are eigenvalues of $\frac 1 \gamma \mathcal{L}$ and it is evident that all of them are located on the boundary of unit disk  $\mathbb{D}(-1,1)$. 
Having the freedom to control $\phi_k$, any point on the boundary can serve as an eigenvalue of some operator $\frac{1}{\gamma} \mathcal{L}$.

The above class of generators, Eq.~(\ref{E2}), has the following property: $\mathcal{L} + \gamma\, {\rm id}$ is completely positive. Is it always possible, for  a given $\mathcal{L}$,  find $\kappa > 0$ such that the map $\mathcal{L} + \kappa\, {\rm id}$ is completely positive? If yes, then one could proceed as in the classical case and define
\begin{equation}\label{}
  \mathcal{L} = (\mathcal{L} + \kappa\, {\rm id}) - \kappa\, {\rm id} = \kappa \Big( \kappa^{-1} \Psi - {\rm id} \Big) ,
\end{equation}
where $\Psi= \mathcal{L} + \kappa\, {\rm id}$ and $\kappa^{-1} \Psi$ is CPTP. Then the spectrum of $\mathcal{L}/\kappa$ would be, therefore,  confined in the unit disk $\mathbb{D}(-1,1)$. However, in the quantum case, this idea does not generally work.

\begin{Example} \label{EX1} Consider a qubit amplitude damping evolution governed by

\begin{equation}\label{}
  \mathcal{L}(\rho) = \gamma \Big(\sigma_- \rho \sigma_+ - \frac 12 \{ \sigma_+\sigma_-,\rho\} \Big) ,
\end{equation}
with $\sigma_\pm = \frac 12 (\sigma_x \pm i \sigma_y)$ being lowering and raising qubit operators. One easily finds the corresponding Choi matrix

\begin{equation}\label{}
 \mathbf{C} = \sum_{i,j=1}^2 |i\>\<j| \otimes \Big( \mathcal{L}(|i\>\<j|) + \kappa\, |i\>\<j| \Big)  = \left( \begin{array}{cc|cc} \kappa & 0 & 0 & -\gamma/2 + \kappa \\ 0 & 0 & 0 & 0 \\ \hline  0 & 0 & \gamma & 0 \\ -\gamma/2 + \kappa & 0 & 0 & - \gamma + \kappa \end{array} \right) .
\end{equation}
One has $\mathbf{C} \geq 0$ if and only if $\kappa \geq \gamma$ and the following $2\times 2$ sub-matrix is semi-positive definite,

\begin{equation}\label{}
 \mathbf{C}_2 =  \begin{pmatrix}
    \kappa &  -\gamma/2 + \kappa \\
     -\gamma/2 + \kappa &  -\gamma + \kappa
  \end{pmatrix} \geq 0 .
\end{equation}
However its determinant ${\rm det}\, \mathbf{C}_2 = -\gamma^2/4< 0$, and hence it is never positive definite which means that $\mathcal{L} + \kappa \, {\rm id}$ is never completely positive. Interestingly, from the structure of $\mathbf{C}$ it follows that the map $\mathcal{L} + \kappa \, {\rm id}$ is even never positive.
\end{Example}

Let us now consider a general purely dissipative GKLS generator,
\begin{equation}\label{}
  \mathcal{L}(\rho) = \Phi(\rho) - \frac 12 \{ \Phi^\ddag(\oper),\rho\} ,
\end{equation}
where $\Phi^\ddag$ denotes a map dual to $\Phi$  w.r.t. the Hilbert-Schmidt inner product in ${\mathcal M}_N(\mathbb{C})$, that is, $(X,\Phi(Y))_{\rm HS} = (\Phi^\ddag(X),Y)_{\rm HS})$, where $(X,Y)_{\rm HS} = {\rm Tr}(X^\dagger Y)$. 

Suppose that the spectrum of $\mathcal{L}$ is known $\mathcal{L}(X_\alpha) = \ell_\alpha X_\alpha$, where $\ell_\alpha = 0$ or ${\rm Re}\,\ell_\alpha < 0$. 

\begin{Proposition}   \label{PRO-R}
The spectrum of $\frac 1R \mathcal{L}$, where

\begin{equation}   \label{R}
    R:= \frac 12 \max_{\lambda_\alpha \neq 0} \frac{|\ell_\alpha|^2}{|{\rm Re}\,\ell_\alpha|} , 
\end{equation}
belongs to $\mathbb{D}(-1,1)$. 
\end{Proposition}
Indeed, in order for the spectrum of the rescaled generator $\ell_\alpha/R$ to belong to $\mathbb{D}(-1,1)$ one requires for all $\ell_\alpha$

\begin{equation}
    |\ell_\alpha/R +1| \leq 1
\end{equation}
and hence $R \geq \frac 12 \frac{|\ell_\alpha|^2}{|{\rm Re}\,\ell_\alpha|}$. Therefore, the minimal scaling which does guarantee that the rescaled spectrum belongs to $\mathbb{D}(-1,1)$ is defined by (\ref{R}). 

We stress that to find $R$, one needs to known the full spectrum of the generator, which is  generally  out of reach. Now, we propose three scaling parameters $\nu_{\text{q},k}$ ($k=1,2,3$) which do not require the knowledge of the spectrum of $\mathcal{L}$. Since $R$ defines a tight scaling,

\begin{eqnarray}
    R \leq \nu_{\text{q},k} , 
\end{eqnarray}
for all $k$.

The authors of Ref.~\cite{Wolf-Cirac} proved the following

\begin{figure}[t]
\begin{center}
\includegraphics[width=0.4\textwidth]{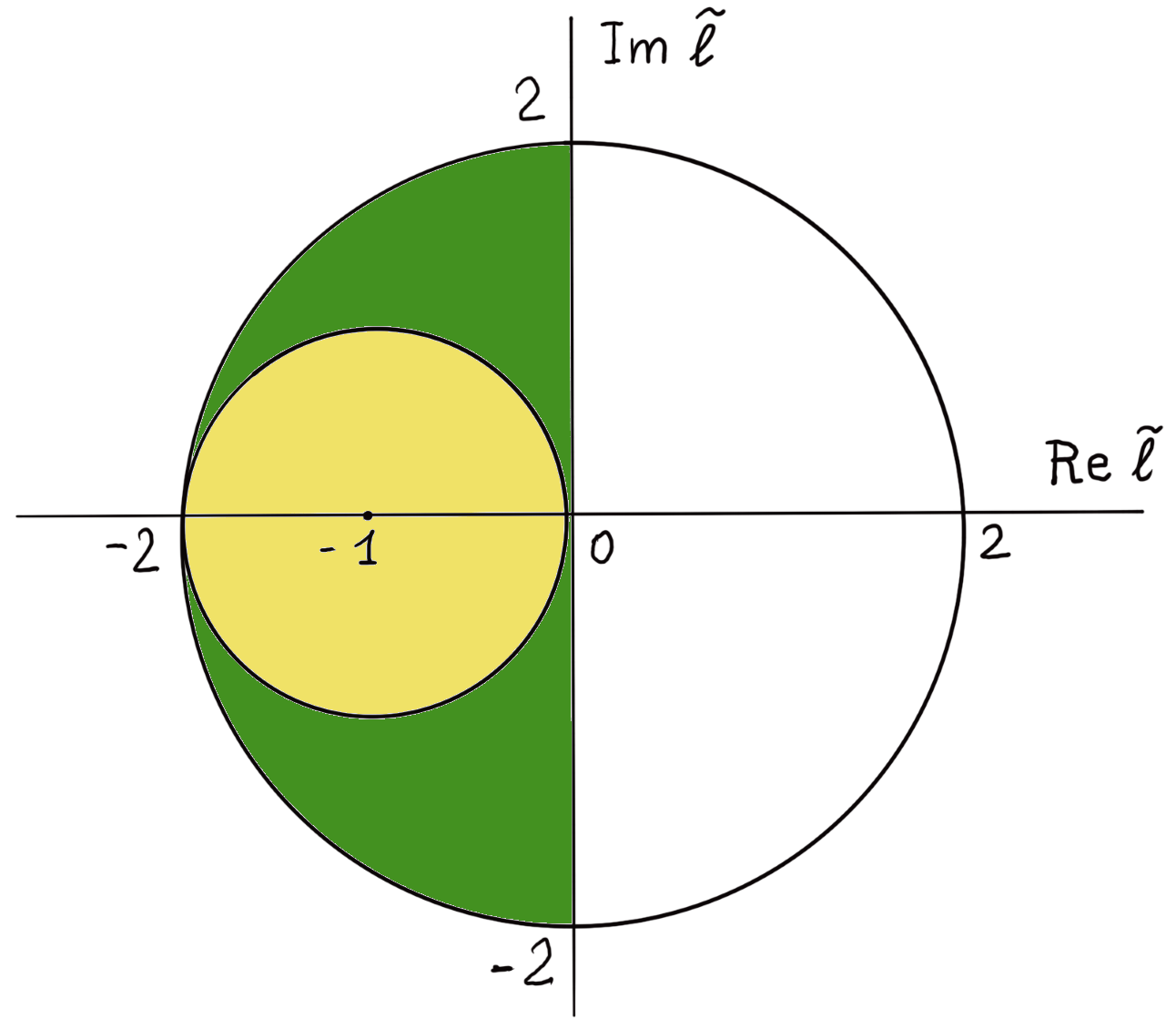}
\caption{According to Theorem 4.1, the spectrum of rescaled dissipative Lindblad operator, $\tilde{\mathcal{L}} = \frac{1}{\nu_{{\rm q},1}} \mathcal{L}$, is confined in 
the disk  $\mathbb{D}(0,2)$. Having maximal eigenvalue $0$, the spectrum is confined to the left semi-disk (green). However, the bound is tighter and the spectrum is confined to the disk $\mathbb{D}(-1,1)$ (yellow).   
\label{F0}}
\end{center}
\end{figure}

\begin{Theorem} Consider the following canonical form of a purely dissipative generator \cite{Erika}

\begin{equation}\label{REP}
    \mathcal{L}(\rho) = \sum_\alpha \gamma_\alpha \left( L_\alpha \rho L_\alpha^\dagger - \frac 12 \{  L_\alpha^\dagger L_\alpha,\rho\} \right) ,
\end{equation}
with  ${\rm Tr}L_\alpha =0$, and ${\rm Tr}  (L_\alpha^\dagger L_\beta) =\delta_{\alpha\beta}$. Then

\begin{equation}\label{norm1}
    \| \mathcal{L} \|_\infty \leq 2 \nu_{{\rm q},1}  \ , \ \ \ \nu_{{\rm q},1} := \sum_\alpha \gamma_\alpha .
\end{equation}
\end{Theorem} 
It is, therefore, clear that the spectrum of $\frac{1}{\nu_{{\rm q},1}} \mathcal{L}$ is confined to $\mathbb{D}(0,2)$. Actually, since the spectrum is located on the left half-plane, it is confined to the left half of the disk; see Fig.~\ref{F0}.

\begin{CON}   \label{CON-0}
    The spectrum of the rescaled generator $\tilde{\mathcal{L}} = \frac{1}{\nu_{{\rm q},1}} \mathcal{L}$ is located in the unit disk   $\mathbb{D}(-1,1) \subset \mathbb{D}(0,2)$; see Fig.~\ref{F0}.
\end{CON}
Since jump operators $L_k$ define an orthonormal basis in $M_N(\mathbf{C})$, we consider three different basis to illustrate our claim. If $L_\alpha = \frac{1}{\sqrt{N}} U_\alpha$, where $U_\alpha$ are unitary operators (e.g., Weyl operators)  satisfying ${\rm Tr}(U_\alpha^\dagger U_\beta)=N \delta_{\alpha\beta}$, we have

\begin{equation}\label{REP-1}
    \mathcal{L}(\rho) = \frac 1N \sum_\alpha \gamma_\alpha \left( U_\alpha \rho U_\alpha^\dagger - \rho \right) ,
\end{equation}
and hence such generator belongs to the family (\ref{E2}). Therefore

\begin{equation}
    \sigma\left( \frac{1}{\nu_{{\rm q},1}} \mathcal{L} \right) \subset \mathbb{D}(-1,1/N), 
\end{equation}
which shows that the scaling parameter $\nu_{{\rm q},1}$ is 
non-optimal. If $L_\alpha = |i\>\<j|$ (i.e., elements of a matrix units basis), then 

\begin{equation}\label{REP-2}
    \mathcal{L}(\rho) = \sum_{i,j=1}^N W_{ij} \left(  |i\>\<j| \rho |j\>\<i| - \frac 12 \{ |j\>\<j|, \rho\} \right) ,
\end{equation}
with $W_{ij}\geq 0$. The spectrum of $\mathcal{L}$ consists of {\em classical} eigenvalues of the corresponding Kolmogorov generator $\mathcal{K}_{ij} = W_{ij} - \delta_{ij} w_j$, with $w_j = \sum_i W_{ij}$, and purely real eigenvalues $\ell_{ij} = - \frac 12 (w_i + w_j)$, (with $i\neq j)$.  We already know that classical eigenvalues belong to $\mathbb{D}(-1,\nu_{\rm c})$, where $\nu_{\rm c} = \max_i w_i$. Note, that $|\ell_{ij}| = \frac 12 (w_i + w_j) \leq \nu_{\rm c}$. Hence, the spectrum of $\frac{1}{\nu_{\rm c}} \mathcal{L}$ belongs to $\mathbb{D}(-1,1)$. It is not unexpected  since Eq.~(\ref{REP-2}) encodes a classical Kolmogorov generator, that is, $\mathcal{K}_{ij} = \<i| \mathcal{L}(|j\>\<j|)|i\>$. We find for the scaling parameter

\begin{equation}
    \nu_{{\rm q},1} =\sum_{i,j=1}^N W_{ij} = \sum_{j=1}^N w_i \geq \max_i w_i = \nu_{\rm c} ,
\end{equation}
and hence it is evident that the spectrum of $\frac{1}{\nu_{{\rm q},1}} \mathcal{L}$ belongs to $\mathbb{D}(-1,1)$.

Finally, we consider an orthonormal basis consisting of generalized  Gell-Mann matrices $F^\dagger_\alpha= F_\alpha$ of order $N$~\cite{ALICKI},

\begin{equation}\label{REP-3}
    \mathcal{L}(\rho) = \sum_\alpha \gamma_\alpha \left( F_\alpha \rho F_\alpha - \frac 12 \{  F_\alpha^2,\rho\} \right) .
\end{equation}
In this case the generator is self-dual and hence the spectrum is purely real,

$$  \sigma(\mathcal{L}) = \{ \ell_0=0 \geq \ell_1 \geq \ldots \geq \ell_{N^2-1} \} . $$
The authors of Ref.~\cite{Rates} proved the following constraint 

\begin{equation}
    |\ell_{N^2-1}| \leq  \sum_\alpha \gamma_\alpha ,
\end{equation}
and hence in this case one has $  \sigma\left( \frac{1}{\nu_{{\rm q},1}} \mathcal{L} \right) \subset [-1,0] \subset \mathbb{D}(-1,0) $.

The above analysis shows that the scaling $1/\nu_{{\rm q},1}$ is not optimal. To find a tighter delineation, we consider  a purely dissipative Lindblad generator

\begin{equation}\label{}
    \mathcal{L}(\rho) = \sum_\alpha \gamma_\alpha \left( L_\alpha \rho L_\alpha^\dagger - \frac 12 \{  L_\alpha^\dagger L_\alpha,\rho\} \right) ,
\end{equation}
which can be compactly rewritten as follows

\begin{equation}\label{}
  \mathcal{L}(\rho) = \Phi(\rho) - \frac 12 \{ \Phi^\ddag(\oper),\rho\} ,
\end{equation}
where $\Phi(\rho) = \sum_\alpha \gamma_\alpha L_\alpha \rho L_\alpha^\dagger$ defines a completely positive map, and 
 $\Phi^\ddag(X) = \sum_\alpha \gamma_\alpha L^\dagger_\alpha X L_\alpha$ and hence $\Phi^\ddag(\oper) = \sum_\alpha \gamma_\alpha L^\dagger_\alpha  L_\alpha$. The map $\Phi$ (being completely positive) has the corresponding Perron-Frobenius eigenvalue $\lambda_{\rm PF}(\Phi)$ such that the spectrum of $\Phi$ is confined to the disc of radius $\rho(\Phi) \equiv \lambda_{\rm PF}(\Phi)$. Let us define a new {\em quantum} scaling parameter 

\begin{equation}
\label{nunu}
 \nu_{\text{q},2} := \| \Phi^\ddag\|_\infty \equiv \| \Phi\|_1 .
\end{equation}
Note that,  due to the Dyo-Russo theorem \cite{Paulsen,Bhatia}, one has $ \| \Phi^\ddag\|_\infty = \|\Phi^\ddag(\oper)\|_\infty $ and $\| \Phi\|_1 = \max_{\|x\|=1} {\rm Tr}\,\Phi(|x\>\<x|)$. 
We propose the following

\begin{CON} \label{CON-1} 
The spectrum of $\frac{ 1}{\nu_{\text{q},2}}\mathcal{L}$ belongs to the unit disk $\mathbb{D}(-1,1)$.  Moreover, $\nu_{\text{q},2}$ provides a tighter delineation than $\nu_{\text{q},1}$, i.e., $\nu_{\text{q},2} \leq \nu_{\text{q},1}$.
\end{CON}

\begin{figure}[t]
\begin{center}
\includegraphics[width=0.55\textwidth]{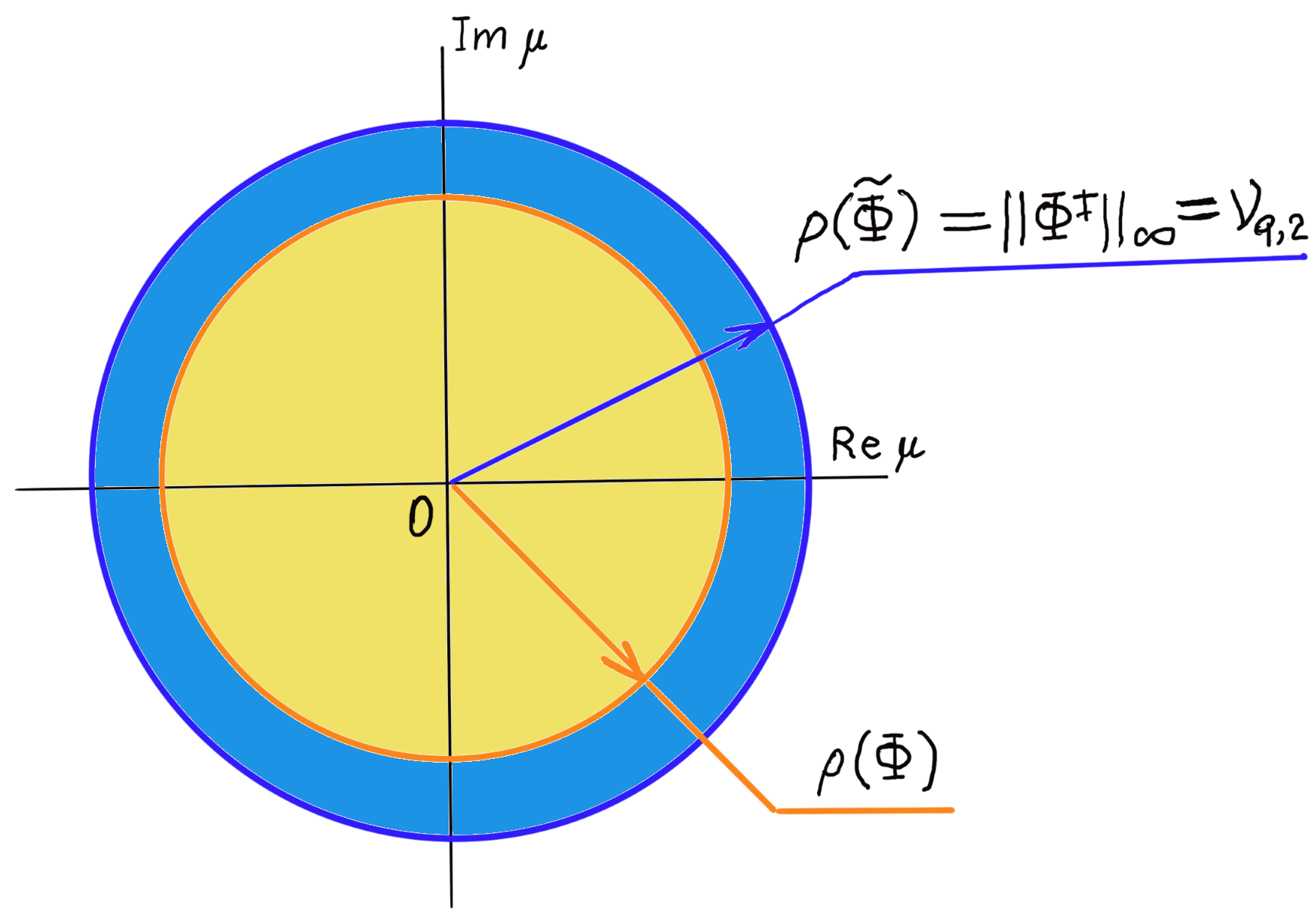}
\caption{The spectrum of the map $\Phi$ is confined within the disk (yellow) of radius $\rho(\Phi)$ (that is the spectral radius of the map). This disk lies strictly inside the disk of radius $\|\Phi^\ddag\|_{\infty}$ (dark blue), while the spectral radius
 $\rho(\widetilde{\Phi})$
of the modified map~(\ref{modified}), attains this value exactly.
\label{F-Ia}}
\end{center}
\end{figure}

Note that
\begin{equation}\label{}
  \mathcal{L}(\rho) = \Phi(\rho) - \frac 12 \{ \Phi^\ddag(\oper),\rho\}  = \left(\Phi(\rho) - \frac 12 \{ \Phi^\ddag(\oper),\rho\} + \nu_{\text{q},2} \, \rho\right) - \nu_{\text{q},2} \, \rho ,
\end{equation}
and hence defining

\begin{equation}\label{D}
  D :=  \Phi^\ddag(\oper)  - \nu_{\text{q},2}\, \oper    ,
\end{equation}
together with
\begin{equation}\label{modified}
  \widetilde{\Phi}(X) = \Phi(X) - \frac 12 (DX + XD)  ,
\end{equation}
one finds

\begin{equation}
    \mathcal{L} = \nu_{\text{q},2} \left(  \frac{1}{\nu_{\text{q},2}}  \widetilde{\Phi} - {\rm id} \right) .
\end{equation}
Observe that in general the new map $\widetilde{\Phi}$ is not completely positive (actually, it is not even positive). Interestingly, $\nu_{\text{q},2}$ serves as a real eigenvalue of  $\widetilde{\Phi}^\ddag$ (and $\widetilde{\Phi}$):

\begin{equation}\label{}
  \widetilde{\Phi}^\ddag(\oper) = \nu_{\text{q},2}\, \oper ,
\end{equation}
which is an analog of (\ref{Wx0}), i.e $\widetilde{W}^{(0)T}\mathbf{x}_0 = \nu_{\rm c}\mathbf{x}_0$. Similarly, if $\rho_{\rm ss}$ is a stationary state of $\mathcal{L}$, i.e. $\mathcal{L}(\rho_{\rm ss})=0$, then

\begin{equation}\label{}
  \widetilde{\Phi}(\rho_{\rm ss}) = \nu_{\text{q},2}\, \rho_{\rm ss} . 
\end{equation}
Note, however, that contrary to the classical case, $\nu_{\text{q},2}$ is not a Perron-Frobenius eigenvalue of $\widetilde{\Phi}$ since $\widetilde{\Phi}$ is not a positive map, and thus the Perron-Frobenius theorem does not apply in this case. It is, therefore, clear that Conjecture~\ref{CON-1} can be equivalently reformulated as follows:  The scaling parameter $\nu_{\text{q},2}$ defines a the spectral radius of $\widetilde{\Phi}$, i.e. $\rho(\widetilde{\Phi})=\nu_{\text{q},2}$. One has therefore

\begin{equation}\label{}
   \rho(\widetilde{\Phi}) \equiv \rho(\widetilde{\Phi}^\ddag) = \| {\Phi^\ddag}\|_\infty \equiv \|\Phi\|_1 ,
\end{equation}
see Fig.~\ref{F-Ia}. This way, the original map $\Phi$ and the ``corrected'' map $\widetilde{\Phi}$ satisfy

\begin{equation}
     \rho({\Phi})   \leq  \| {\Phi^\ddag}\|_\infty =      \rho(\widetilde{\Phi})  \leq \| {\widetilde{\Phi}^\ddag}\|_\infty .
\end{equation}

Interestingly, Conjecture \ref{CON-1} holds for important classes of generators. Below we present a few illustrative examples. 

\begin{Example} Consider the generator from Example \ref{EX1}, 

\begin{equation}\label{AD}
  \mathcal{L}(\rho) = \gamma \Big(\sigma_- \rho \sigma_+ - \frac 12 \{ \sigma_+\sigma_-,\rho\} \Big) .
\end{equation}
One finds

\begin{equation}
  \nu_{\text{q},2} =  \| \Phi^\ddag(\oper)\|_\infty = \gamma \|\sigma_+  \sigma_-\|_\infty = \gamma  \ , \ \ \  \nu_{\text{q},1} = \gamma ,
\end{equation}
and hence $ \nu_{\text{q},2} =  \nu_{\text{q},1}$. 
Note that

$$   \mathcal{L}(\sigma_\pm) = -\frac \gamma 2\sigma_\pm  \ , \ \
\mathcal{L}(\sigma_z) = - \gamma \sigma_z , $$
which implies  that the spectrum of $\frac {1}{\nu_{\text{q},2}} \mathcal{L}$ reads $\{0,-\frac 12 ,-\frac 12, -1\}$. 

\end{Example}

\begin{Example}   \label{EX2}

Consider now pure dephasing evolution governed by the following generator,

\begin{equation}
    \mathcal{L}(\rho) = -[A,[A,\rho]] ,
\end{equation}
with an arbitrary Hermitian $N \times N$ matrix $A$. It can be rewritten in the standard Lindblad form as follows,

\begin{equation}
    \mathcal{L}(\rho) = 2 A\rho A - A^2\rho - \rho A^2 .
\end{equation}
Note that $\mathcal{L}^\ddag = \mathcal{L}$. In the qubit case, taking $A = \frac{\gamma}{\sqrt{2}}\sigma_z$, we recover the qubit dephasing generator
\begin{equation}   \label{szsz}
     \mathcal{L}(\rho) = \gamma( \sigma_z \rho\sigma_z - \rho) .
\end{equation}
In this case case we find

\begin{equation}
 \nu_{\text{q},2} = \gamma \ , \ \ \ \nu_{\text{q},1} = 2\gamma ,   
\end{equation}
and the spectrum of $\frac{1}{\nu_{\text{q},2}}   \mathcal{L}$ reads $\{0,0,-2,-2\}$. Hence, in this case the spectrum belongs to the boundary of the disk $\mathbb{D}(-1,1)$ showing that the scaling $\nu_{\text{q},2}$, contrary to $\nu_{\text{q},1}$, is tight for this model.  In the general case let $A e_k = a_k e_k$ and let $|a_1| \geq |a_2|\geq \ldots\geq |a_n|$.  Hence

$$  \mathcal{L}(|e_k\>\<e_\ell|) = - (a_k - a_l)^2 |e_k\>\<e_\ell| \  , \ \ \ k,\ell=1,\ldots,N .  $$
One finds for the scaling parameters

$$  \nu_{\text{q},2} = \| \Phi\|_\infty = \|\Phi(\oper)\|_\infty = 2 a_1 ^2 \ , \ \ \ \nu_{\text{q},1} = 2 \sum_i a_i^2 . $$
The spectrum of $\frac{1}{\nu_{\text{q},2}} \mathcal{L} $ reads

$$    \lambda_{k\ell} = - \frac{1}{2} ( \tilde{a}_k - \tilde{a}_\ell)^2 , \ \ \  \tilde{a}_k = \frac{a_k}{|a_1|} .$$
and hence $\lambda_{k\ell} \in [-2,0]$ which supports the Conjecture.

\end{Example}

\begin{Example} \label{EX-ggg} The qubit generator (\ref{szsz}) may be generalized as follows,

\begin{equation}   \label{Pauli}
     \mathcal{L}(\rho) = \sum_{k=1}^3 \gamma_k( \sigma_k \rho\sigma_k - \rho) ,
\end{equation}
with $\gamma_k \geq 0$. Again, one has $ \mathcal{L}^\ddag =  \mathcal{L}$. The map $\Phi$ reads $\Phi(\rho) =  \sum_k \gamma_k \sigma_k \rho\sigma_k$ and hence

$$  \nu_{\text{q},2}=   \|\Phi\|_\infty = \gamma_1 + \gamma_2 + \gamma_3  \ , \ \ \ \nu_{\text{q},1}= 2(\gamma_1 + \gamma_2 + \gamma_3) . $$
The spectrum of (\ref{Pauli}) reads $\mathcal{L}(\sigma_k) = \lambda_k \sigma_k $, where

$$ \lambda_1 = -2(\gamma_2 + \gamma_3) \ , \ \ \lambda_2 = -2(\gamma_3 + \gamma_1) \ , \ \ \lambda_3 = -2(\gamma_1 +\gamma_2) \ ,$$
and hence the spectrum of $\frac{1}{ \nu_{\text{q},2}} \mathcal{L}$ belongs to $\left[\frac{|\lambda_{\rm min}|}{\nu_{\text{q},2}},0\right] \subset [-2,0]$, where $\lambda_{\rm min}= \min_k \lambda_k$.

\end{Example}
 Finally, let us compare $\nu_{\text{q},1}$ and $\nu_{\text{q},2}$ for three Lindblad operators represented by Eqs.~(\ref{REP-1})--(\ref{REP-3}).  For $\mathcal{L}$ defined in Eq.~(\ref{REP-1}), we have

\begin{equation}
    \nu_{\text{q},1} = \sum_{i,j} W_{ij} \geq \nu_{\text{q},2} = \max_i w_i .
\end{equation}
For Eq.~(\ref{REP-2}),  we have $ \nu_{\text{q},1} = \nu_{\text{q},2} = \sum_{\alpha}\gamma_\alpha$. 
Finally, for operator~(\ref{REP-3}), 

\begin{equation}
  \nu_{\text{q},2} = \|\sum_\alpha \gamma_\alpha F^2_\alpha\|_\infty \leq \sum_{\alpha}\gamma_\alpha  =  \nu_{\text{q},1} , 
\end{equation}
due to $\|F_\alpha^2\|_\infty < {\rm Tr}F^2_\alpha =1$. Hence, $\nu_{\text{q},1} \geq \nu_{\text{q},2}$ which indeed shows that $\nu_{\text{q},2}$ provides a tighter delineation than $\nu_{\text{q},1}$.

In Appendix \ref{APP} we provide the detailed analysis of a class of covariant Lindblad operators acting on a system of any
dimension $N$. 

\section{Beyond complete positivity} 
\label{Beyond}

Now we are going to relax the requirement of complete positivity and consider semi-groups of positive and trace-preserving (PTP) maps. Recall that both CPTP and PTP maps have the same spectral properties. Indeed, due the  quantum version of the corresponding Perron-Frobenius theorem \cite{Perron-1,Perron-2,WOLF}, spectra of such maps are confined to the unit disk in the complex plane and the Perron-Frobenius eigenvalue equals 1. The structure of generators giving rise to PTP semigroups is characterized as follows~\cite{KOS-72}:

\begin{Proposition} $\mathcal{L}$ is a generator of PTP semigroups if and only if for any orthonormal basis $\{|1\>,\ldots,|n\>\}$ the following $N \times N$ matrix

\begin{equation}   \label{L-K}
    {\cal K}_{ij} := \<i|\mathcal{L}(|j\>\<j|)|i\> ,
\end{equation}
defines a classical generator, i.e., ${\cal K}_{ij} \geq 0$ for $i\neq j$, together with the normalization condition $\sum_j {\cal K}_{ij}=0$.
\end{Proposition}
Any generator of a PTP semi-group can be represented as
\begin{equation}\label{GKLS-P}
  \mathcal{L}(\rho) = -i[H,\rho] + \sum_k \gamma_k \Big( L_k \rho L_k^\dagger - \frac 12 \{ L_k^\dagger L_k, \rho\} \Big) ,
\end{equation}
but now some of the rates $\gamma_k$ may be negative. Evidently,  Lindblad operators  belong to this class.   
The above characterization suggests another definition of a scaling parameter. 


\begin{CON} \label{CON-3} If $\mathcal{L}$ is a generator of positive trace-preserving dynamics and 

\begin{equation}  \label{kappa}
    \nu_{\text{q},3} := \max_{\|\psi\|=1} | \< \psi|\mathcal{L}(|\psi\>\<\psi|)|\psi\>| ,
\end{equation}
then the spectrum of $\frac{1}{ \nu_{\text{q},3}}  \mathcal{L} $ is confined in the unit disk $\mathbb{D}(-1,1)$.   
\end{CON}
This means that, for any orthonormal basis $\{|1\>,\ldots,|N\>\}$, we can define ${\cal K}_{ij}$ via Eq.~(\ref{L-K}) and then calculate the classical rescaling parameter $\nu_{\text{c}}$. Finally, $ \nu_{\text{q},3}$ is nothing more than $\nu_{\text{c}}$ optimized for all orthonormal bases. Moreover, we conjecture that $ \nu_{\text{q},3}$ defined in Eq.~(\ref{kappa}) is {\em optimal}, i.e., minimal, scaling parameter such that the spectrum of the corresponding rescaled generator belongs to $\mathbb{D}(-1,1)$.  It is clear, though, that determining $\nu_{\text{q},3}$ is significantly more demanding than computing $\nu_{\text{q},2} = \|\Phi^\ddag(\oper)\|_\infty$, { as it involves a maximization over the sphere $\|\psi\| = 1$ in $\mathbb{C}^N$. Such optimization is non-convex and non-linear and therefore hard~\cite{Absil}. It is an interesting question whether it can be cast as a semidefinite program (SDP) and thus made computationally tractable~\cite{Boyd}.}

\begin{Example}
For the Lindblad generator defined in Eq.~(\ref{AD}) which involves only a single rate $\gamma >0$, it is easy to find that 

\begin{equation}
    \nu_{\text{q},1}= \nu_{\text{q},2}= \nu_{\text{q},3}= \gamma .
\end{equation}
Consider now a Lindblad generator from Example~\ref{EX-ggg}.
Recall that $\nu_{\text{q}_1} = 2(\gamma_1 + \gamma_2 + \gamma_3)$ and $\nu_{\text{q},2} = \frac 12 \nu_{\text{q}_1}$.  Interestingly, the $\nu_{\text{q},3}$ parameter reads as follows

\begin{equation}
    \nu_{\text{q},3} = \nu_{\text{q},2} - \min\{\gamma_1,\gamma_2,\gamma_3\}  \leq \nu_{\text{q},2} . 
\end{equation}
Assuming $\gamma_1 \geq \gamma_2 \geq \gamma_3\geq 0$ one finds  $\nu_{\text{q,3}} =  \gamma_1 + \gamma_2$ and hence

\begin{equation}  \label{nnn}
     \nu_{\text{q},3} \leq  \nu_{\text{q},2} \leq  \nu_{\text{q},1} .
\end{equation}
Note that if $\gamma_3=0$, then $\nu_{\text{q},3} = \nu_{\text{q},2}$. The spectrum of rescaled generator $\frac{1}{\nu_{\text{q},3}} \mathcal{L}$ reads: 

$$   \left\{ -2, - 2\frac{\gamma_2 + \gamma_3}{\gamma_1 + \gamma_2},- 2\frac{\gamma_3 + \gamma_1}{\gamma_1 + \gamma_2}, 0 \right\} ,  $$
and hence it is clear that $\nu_{\text{q},3}$ is an optimal scaling parameter. 

In fact, the same applies to a generator giving rise to PTP dynamics. In this case, $\gamma_3$ can be negative, provided that $\gamma_1 + \gamma_3 \geq 0$ and $\gamma_2 + \gamma_3 \geq 0$ \cite{PR2022}. Again, one finds $\nu_{\text{q},3} = \gamma_1 + \gamma_2$, but one cannot define $\nu_{\text{q}}$ via $\|\Phi(\oper)\|_\infty$, since the map $\Phi(\rho) = \sum_k \gamma_k \sigma_k \rho \sigma_k$ is not even positive in this case. 

Consider, for example, $\gamma_1 = \gamma_2 = 1$ and $\gamma_3 = -1$ leading to $\nu_{\text{q},3} = 2$. The spectrum of the rescaled generator then reads $\{-2, 0, 0, 0\}$, that is, only points from the boundary of the disk $\mathbb{D}(-1,1)$ belong to the spectrum. Once again, it is evident that $\nu_{\text{q},3} = 2$ cannot be improved.  Interestingly, for the generator defined in (\ref{Pauli}) one has

\begin{equation}
    \nu_{\text{q},3} = R ,
\end{equation}
with $R$ defined in (\ref{R}), and hence $\nu_{\text{q},3}$ is already tight. 

\end{Example}

\begin{Example}
    To illustrate optimality of  $\nu_{\text{q},3}$ consider the following generator $\mathcal{L}(\rho) = A \rho A - \frac 12 \{A^2,\rho\}$, with Hermitian traceless $A$. 
One finds 
\begin{equation}
    \nu_{\text{q},3}  = \max_{\|\psi\|=1} \left( \<A^2\>_\psi - \<A\>^2_\psi \right) ,
\end{equation}
where $\<A\>_\psi = \<\psi|A|\psi\>$.   To simplify the analysis consider $A = {\rm Diag}[a_1,\ldots,a_N]$, with $a_1\geq a_2 \geq \ldots \geq a_N$.  Now, since $A$ is traceless $a_1 > 0 > a_N$. One finds

$$ \nu_{\text{q},3} = \frac 14 (a_1-a_N)^2= \frac 14 (a_1+|a_N|)^2 , $$
and the maximum is realized for $\psi = \left( \frac{1}{\sqrt{2}},0,\ldots,0,\frac{1}{\sqrt{2}}\right)$. The spectrum of $\mathcal{L}$ consists of real eigenvalues $\ell_{ij} = -\frac 12 (a_i-a_j)^2$. Hence, 

$$  |\ell_{1N}| =  \max_{i\neq j}|\ell_{ij}| =  2\nu_{\text{q},3} , $$
which shows that $\ell_{1N}/\nu_{\text{q},3} = -2$ belongs to the boundary of the disk $\mathbb{D}(-1,1)$. Clearly, $ \nu_{\text{q},3} = R$. 
    
\end{Example}

\section{Time dependent generators: beyond Markovian regime}
\label{Sec-NONM}

Let us briefly analyze what happens if the generator does depend on time. In particular we address the question: Is it possible to find an appropriate  time dependent rescaling parameter in the non-Markovian regime? In the classical case the corresponding master equation reads
\begin{equation}   \label{SKS}
    \dot{S}(t) = \mathcal{K}(t) S(t) \ ,
\end{equation}
and its solution has the following form:
\begin{equation}   \label{S(t)}
    S(t) = \mathcal{T} \exp\left( \int_0^t \mathcal{K}(\tau) d\tau \right) ,
\end{equation}
where $\mathcal{T}$ is  the time ordering operation. Now, if the time-dependent generator $\mathcal{K}(t)$ satisfies the Kolmogorov conditions, i.e., $\mathcal{K}_{ij}(t) \geq 0$ for $i \neq j$ and $\sum_i \mathcal{K}_{ij}(t) = 0$, then Eq.~(\ref{S(t)}) defines a legitimate classical dynamical map; that is, $S(t)$ is a stochastic matrix for all $t \geq 0$.  Note that in this case, the family $\{S(t)\}_{t \geq 0}$ enjoys the  divisibility property:

\begin{equation}  \label{SSS}
    S(t) = S(t,u) S(u)  , 
\end{equation}
for any $t\geq u \geq 0$, and $S(t,u)$ is a stochastic matrix defined by 

\begin{equation}
     S(t,u) =  S(t) S^{-1}(u) = \mathcal{T} \exp\left( \int_u^t \mathcal{K}(\tau) d\tau \right) .
\end{equation}
Such divisible dynamical maps represent classical Markovian evolution  \cite{Piilo-class,PR2022}. It is therefore clear that for Markovian evolution the spectrum of rescaled generator $\mathcal{K}(t)/{\nu_{\rm c}(t)} $ belongs to $\mathbb{D}(-1,1)$, where 
\begin{equation}
    \nu_{\rm c}(t) = \max_i |\mathcal{K}_{ii}(t)| . 
\end{equation}
Now we show that for non-Markovian evolution, i.e. when Kolmogorov conditions $\mathcal{K}_{ij}(t) \geq 0$ (for $i \neq j$) are violated, in general such rescaling is not possible.

\begin{Example} To illustrate the above statement, we consider the evolution of a $N=2$-level system  represented by
\begin{eqnarray}   \label{Sab}
    S(t) = \left( \begin{array}{cc} a(t) & 1- b(t) \\ 1-a(t) & b(t) \end{array} \right) , 
\end{eqnarray}
with $a(t),b(t) \in [0,1]$, and $a(0)=b(0)=1$. Such dynamics is governed by the following time-dependent generator \cite{Piilo-class,PR2022}

\begin{equation}
    \mathcal{K}(t) = \dot{S}(t) S^{-1}(t) =  \left( \begin{array}{cc} -k_1(t)  & k_2(t) \\ k_1(t) & -k_2(t) \end{array} \right)
\end{equation}
where

$$ k_1(t) = -\frac{w(t) + \dot{b}(t) }{a(t) + b(t) -1} \ , \ \ \ k_2(t) = \frac{w(t) - \dot{a}(t)}{a(t) + b(t) -1} ,$$
and $w(t) = \dot{a}(t) b(t) - \dot{b}(t)a(t)$. One obviously has $\sum_i \mathcal{K}_{ij}(t)=0$. However, the conditions $k_1(t) \geq 0$ and $k_2(t) \geq 0$ are no longer guaranteed. The spectrum of $\mathcal{K}(t)$ consists of $\{0, \mathrm{Tr}\,\mathcal{K}(t)\}$. Hence, whenever $\mathrm{Tr}\,\mathcal{K}(t) > 0$ for some $t > 0$, the scaling does not exist, since the nonzero eigenvalue lies in the right half of the complex plane and therefore cannot belong to the disk $\mathbb{D}(-1,1)$.

Consider, for example,

\begin{equation}
    a(t) = \frac 12 e^{-\frac 12 t} (1+ \cos t)  \  ,\ \ \ \  b(t) = \frac 12 (1+ e^{-4 t}) . 
\end{equation}
It is easy to see that both $k_1(t)$ and $k_2(t)$ are temporally negative showing that Eq.~(\ref{Sab}) represents a non-Markovian classical dynamics. Similarly, 

$$ {\rm Tr}\mathcal{K}(t) = -k_1(t) - k_2(t) = \frac{\dot{a}(t) + \dot{b}(t)}{a(t) + b(t) -1} , $$ 
is temporally strictly positive. 


\end{Example}

Consider now the quantum evolution governed by $    \dot{\Lambda}_t = \mathcal{L}_t \Lambda_t$, 
where the time-dependent Lindblad operator $\mathcal{L}$ has the following form:

\begin{equation}
    \mathcal{L}(\rho) = - i[H(t),\rho] + \sum_\alpha \gamma_\alpha(t) \left( L_\alpha(t) \rho L_\alpha(t) - \frac 12 \{ L_\alpha^\dagger(t) L_\alpha(t),\rho\} \right) ,
\end{equation}
with time dependent Hamiltonian $H(t)$, rates $\gamma_\alpha(t)$, and jump operators $L_\alpha(t)$. If all rates $\gamma_\alpha(t) \geq 0$, then

\begin{equation}   \label{}
    \Lambda_t = \mathcal{T} \exp\left( \int_0^t \mathcal{L}_\tau d\tau \right) ,
\end{equation}
defines a legitimate quantum dynamical map, i.e. $\Lambda_t$ is CPTP for all $t \geq 0$. Moreover, in this case $\Lambda_t$ enjoys the following composition law [an analog of (\ref{SSS})]:

\begin{equation}
    \Lambda_t = \Lambda_{t,u} \Lambda_u \ ,
\end{equation}
for $t \geq u \geq 0$, and the intermediate map (a propagator) 

\begin{equation}   \label{}
    \Lambda_{t,u} = \Lambda_t \Lambda_u^{-1} = \mathcal{T} \exp\left( \int_u^t \mathcal{L}_\tau d\tau \right) ,
\end{equation}
is CPTP. Such evolution is called CP-divisible (propagators are CP). One calls a quantum dynamical map $\{\Lambda_t\}_{t\geq 0}$ to be Markovian if it is CP-divisible; see recent reviews~\cite{NM1,NM2,NM3,NM4,PR2022}.  $\{\Lambda_t\}_{t\geq 0}$ is called P-divisible if all propagators $\Lambda_{t,u}$ are positive and trace-preserving (for the whole hierarchy of divisible maps see, e.g., Ref.~\cite{Sabrina}). Now, the map is P-divisible if for any orthonormal basis $\{|1\rangle,\ldots,|N\rangle\}$

\begin{equation}   \label{L-K-t}
    {\cal K}_{ij}(t) := \<i|\mathcal{L}_t(|j\>\<j|)|i\> ,
\end{equation}
defines a classical time-dependent Markovian generator, i.e.,  $ {\cal K}_{ij}(t) \geq 0$ for $i \neq j$.

\begin{Cor} If $\mathcal{L}_t$ generates a P-divisible dynamical map $\{\Lambda_t\}_{t\geq 0}$, then 
the spectrum of the rescaled generator $\mathcal{L}_t/\nu_{\text{q},3}(t)$, with

\begin{equation}  \label{}
    \nu_{\text{q},3}(t) := \max_{\|\psi\|=1} | \< \psi|\mathcal{L}_t(|\psi\>\<\psi|)|\psi\>| ,
\end{equation}
belongs to $\mathbb{D}(-1,1)$. 
\end{Cor}
However, for non-Markovian dynamics that are not P-divisible, such time-dependent rescaling is not guaranteed.

\begin{Example} Consider the evolution of $N=2$-level system governed by

\begin{equation}
    \mathcal{L}_t(\rho) = \frac 12 \sum_{k=1}^3 \gamma_k(t)( \sigma_k \rho \sigma_k- \rho) .
\end{equation}
The spectrum of $\mathcal{L}_t$ reads 

$$ \{0,-[\gamma_1(t) + \gamma_2(t)],-[\gamma_2(t) + \gamma_3(t)],-[\gamma_3(t) + \gamma_1(t)]\} , $$
and P-divisibility requires that all eigenvalues are non-positive  \cite{PR2022}, i.e.

$$  \gamma_1(t) + \gamma_2(t) \geq 0 \ , \  \gamma_2(t) + \gamma_3(t) \geq 0 \ , \  \gamma_3(t) + \gamma_1(t) \geq 0 \ ,  $$
for all $t \geq 0$. Let
\begin{equation}
    \gamma_1(t) = \gamma_2(t) = 2 e^{-\frac 12 t}\left( \frac 12 + \cos t\right)    \ , \ \ \gamma_3(t) =  e^{-\frac 38 t} .
\end{equation}
Such time-dependent generator gives rise to well defined CPTP dynamical map \cite{Sabrina-Entropy}.  
Evidently, P-divisibility is violated since $ \gamma_1(t) + \gamma_2(t)=2 \gamma_1(t)$ is temporally negative. 
Now, an eigenvalue `$-2 \gamma_1(t)$' is temporally positive and hence cannot belong to $\mathbb{D}(-1,1)$. 
\end{Example}
These simple examples  show that beyond Markovian regime we cannot, in general, guarantee the existence of a rescaling which can bound the spectrum of a time-dependent generator to $\mathbb{D}(-1,1)$. For related discussions on spectral properties and non-Markovianity, see Ref.~\cite{Sabrina-Chiara}.

\section{On the spectra of random generators}
\label{Sec_typical}

The concept of "random Markov generators" was introduced recently~\cite{timm,Random-1,Random-2,Ca19,COOG19,SRP20} and continues to be actively discussed in different contexts, see, e.g, Refs.~\cite{sa2022,kulkarni2022}. Considered as \textit{typical}, these 'average citizens' of the space of all possible Markov generators exhibit a high degree of universality in their spectral properties~\cite{Random-1,Random-2}.

There is a natural  question about the relationship between the  spectra of the typical generators of Markovian evolution, after the rescaling, and the bounds discussed in the previous two sections. Specifically, how do the spectral densities of  the rescaled random  generators compare to these  bounds, and what insights can be drawn about their asymptotic behavior?

Let us start with the classical case. The notion of a random stochastic matrix $M$ was discussed in Ref.~\cite{S14,BCC12}. Such matrices can be generated  by first filling the entries with independent and identically distributed (i.i.d.) random non-negative numbers, sampled, e.g., from a $\chi^2$-distribution~\cite{Random-2}, and then enforcing column-wise normalization~\cite{S14}. Alternatively, they can be sampled by generating dense random directed graphs and assigning random transition probabilities to the graph edges~\cite{BCC12}.

\begin{figure}[t]
    \centering
    \begin{subfigure}[b]{0.32\textwidth}
        \includegraphics[width=\textwidth]{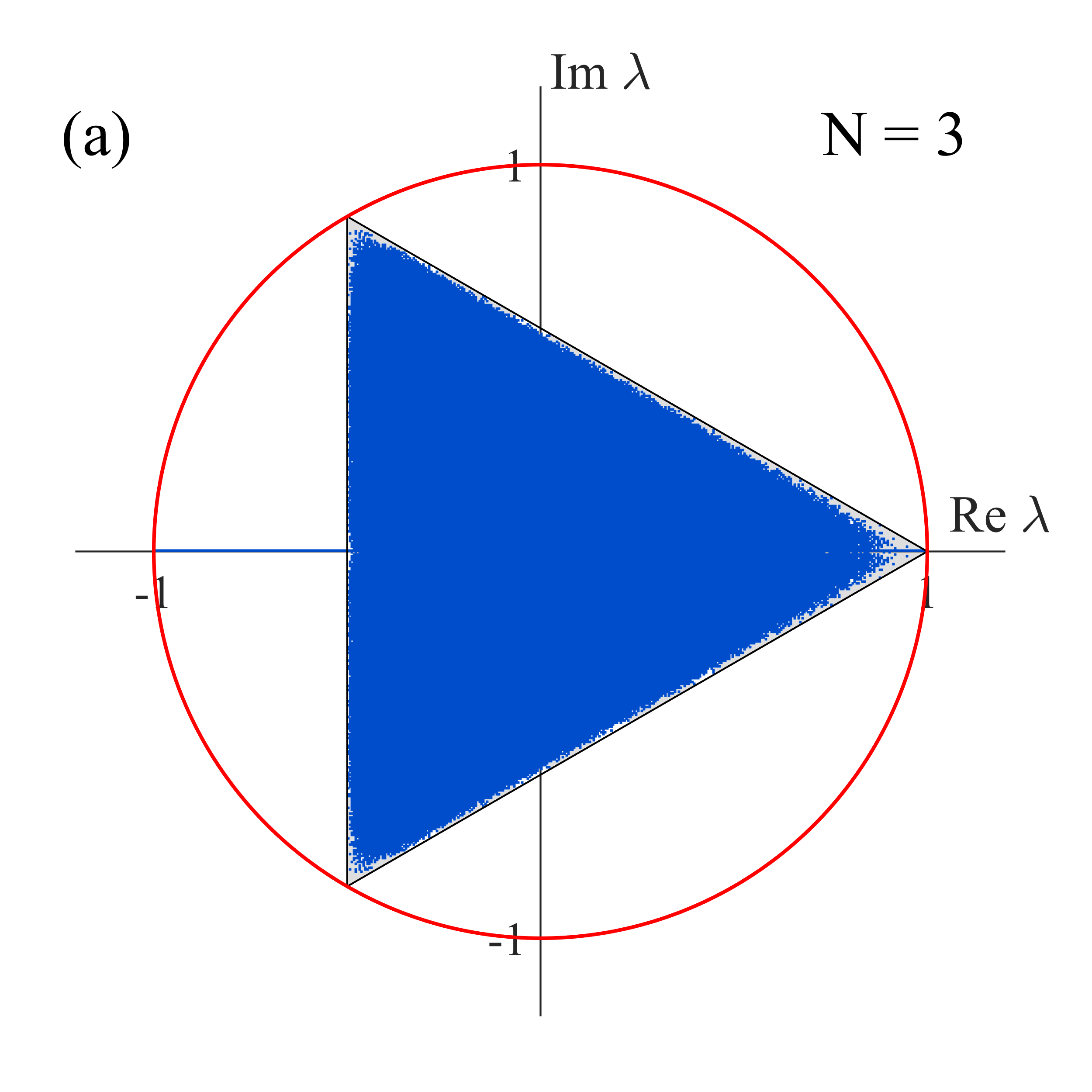}
    \end{subfigure}
    \hfill
    \begin{subfigure}[b]{0.32\textwidth}
        \includegraphics[width=\textwidth]{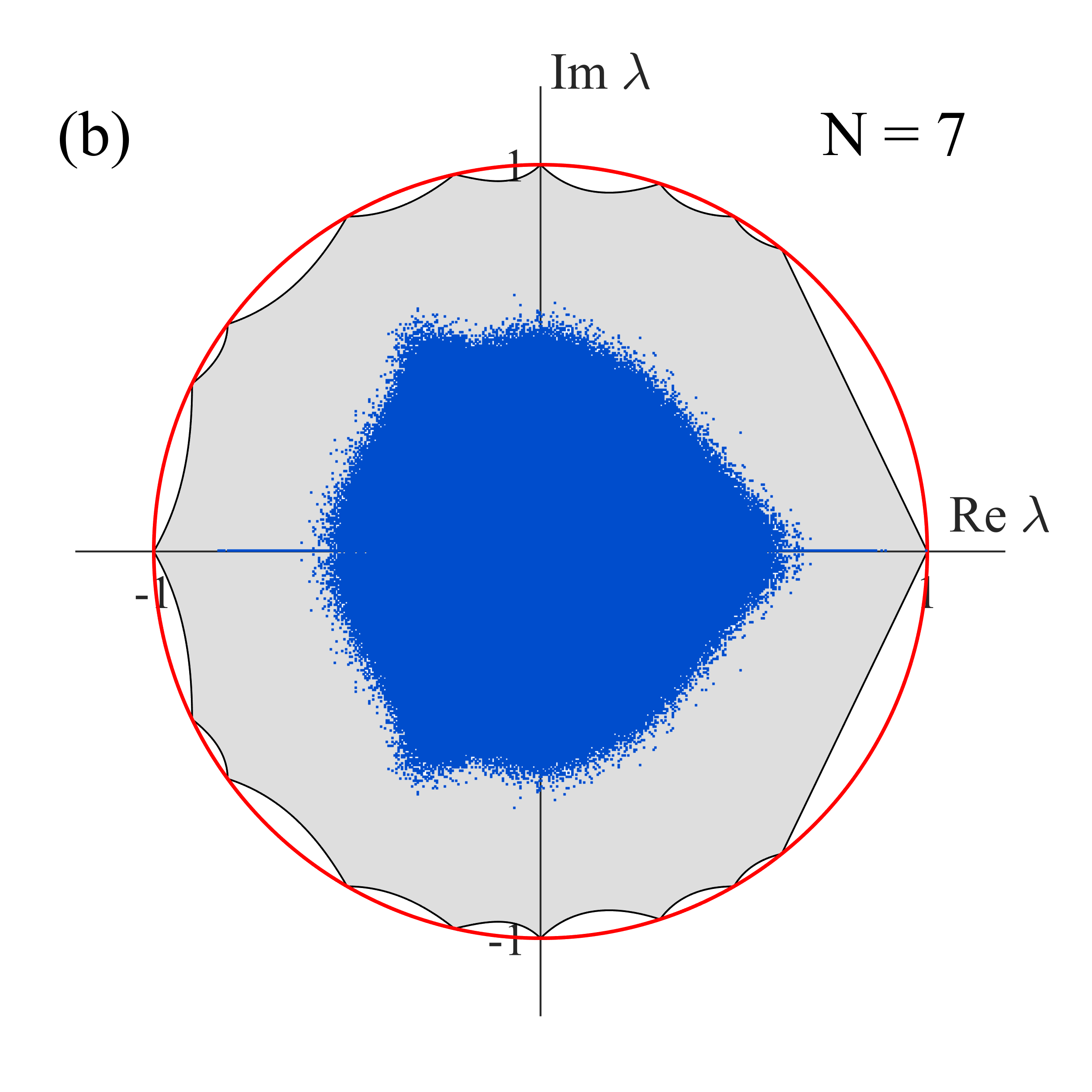}
    \end{subfigure}
    \hfill
    \begin{subfigure}[b]{0.32\textwidth}
        \includegraphics[width=\textwidth]{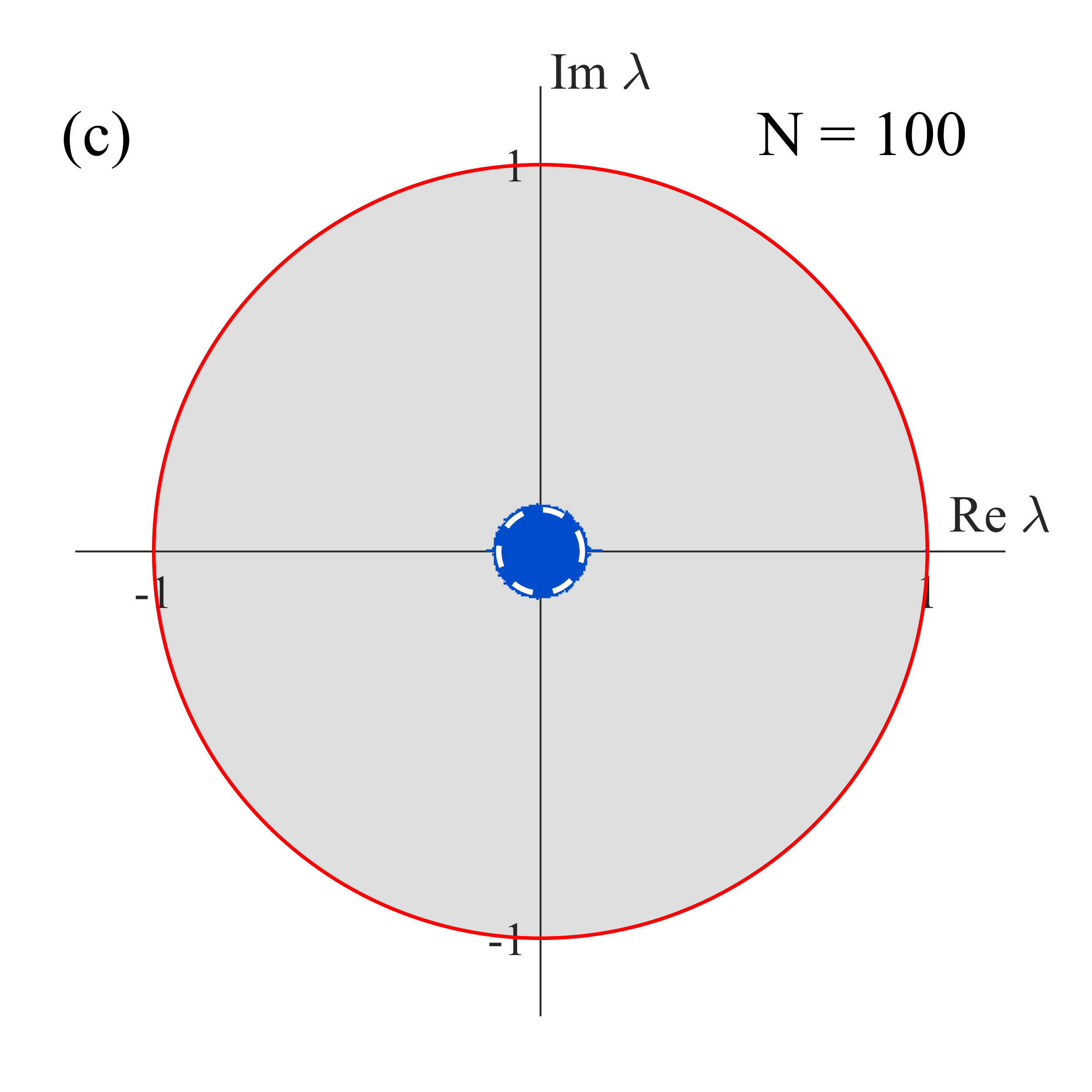}
    \end{subfigure}
    \caption{Spectral support of random stochastic matrices $S$ for different values of $N$. Results of random sampling are presented as $400 \times 400$ histograms (dark blue), where only bins containing at least one eigenvalue are shown. The number of samples is $10^8$, $10^6$, and $10^5$ for $N=3$ (a), $N=7$ (b), and $N=100$ (c), respectively. Karpelevi\u{c} regions $\Theta_N$ are indicated in light gray. The dashed line in panel (c) indicates the circle of radius $0.1$.
\label{F-5} }
\end{figure}

Following the prescription~\cite{S14} for $N=3$, we are able to out-shape Karpelevi\u{c} region $\Theta_3$; see Fig.~\ref{F-5}(a). However, already for $N=7$, even with extensive sampling, $\Theta_7$ remains unresolved; see Fig.~\ref{F-5}(b). For $N=100$, the sampled eigenvalues are strongly localized within a disk of radius $\simeq 0.1$, see Fig.~\ref{F-5}(c), even though the corresponding Karpelevi\u{c} region $\Theta_{100}$ almost coincides with the unit disk.  This trend persists for larger values of $N$, with the spectral support remaining confined within a disk of radius $1/\sqrt{N}$~\cite{S14}.

From the perspective of random matrix theory~\cite{Zeitouni}, the asymptotic localization of the spectrum is a consequence of column-wise normalization: After this, the typical scale of the  matrix entries is of order $1/N$. Hence, when treating the stochastic matrix -- for the bulk of the spectrum -- as a member of the real Ginibre ensemble~\cite{Zeitouni},  we could expect a spectral support confined to the disk of radius $1/\sqrt{N}$.

To escape the 'spell of localization', 
we must use (or construct) \textit{\underline{a}}typical stochastic matrices. This implies, for example, that the corresponding models should respect certain topological constraints, such as, e. g.,  pair-wise neighbor interactions, or/and that transition probabilities should not be independently random, etc. These features are clearly reflected in the structure of  stochastic matrices that resolve the boundaries of the Karpelevi\u{c} regions~\cite{Smigoc-3}.

\begin{figure}[h]
    \centering
    \begin{subfigure}[b]{0.32\textwidth}
        \includegraphics[width=\textwidth]{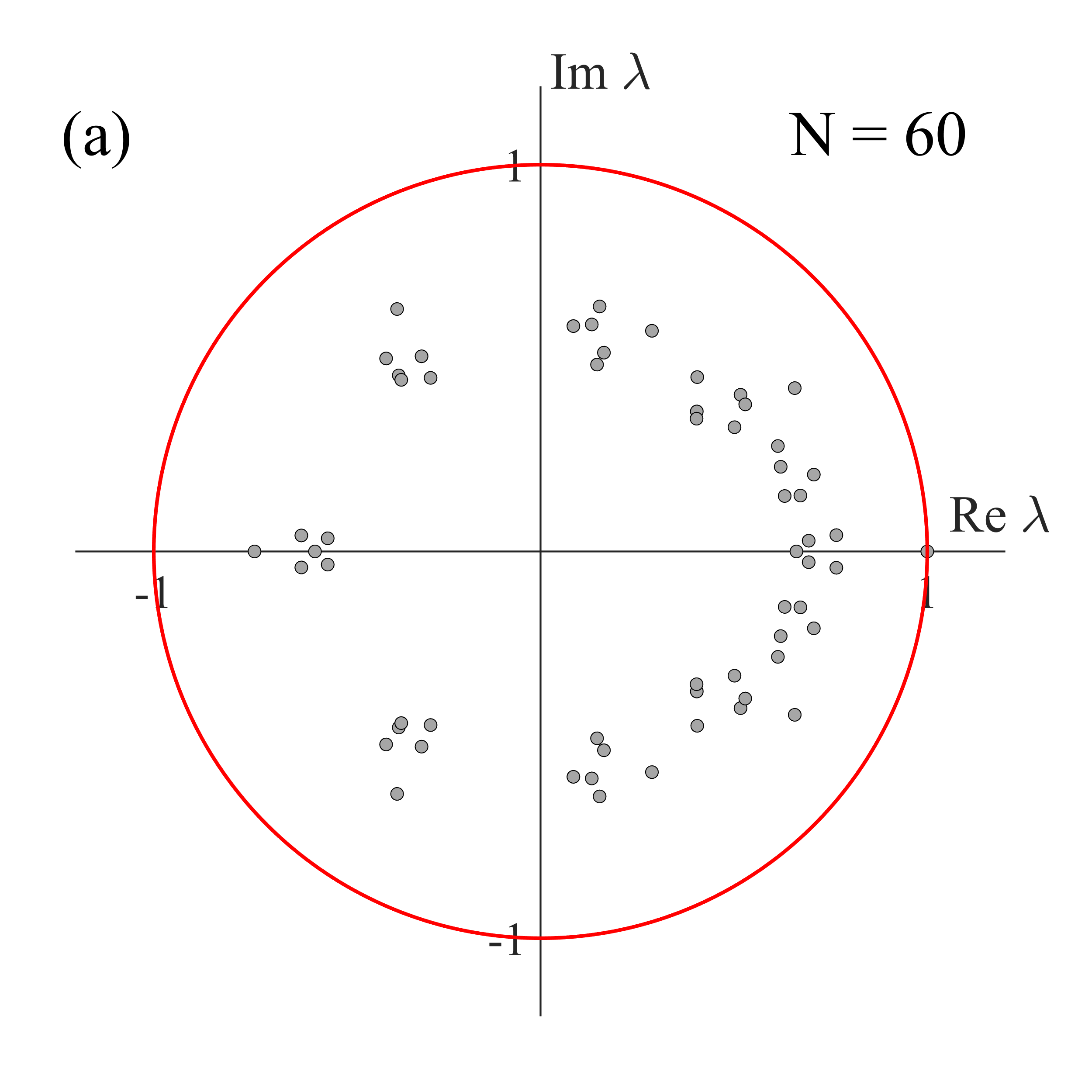}
    \end{subfigure}
    \hfill
    \begin{subfigure}[b]{0.32\textwidth}
        \includegraphics[width=\textwidth]{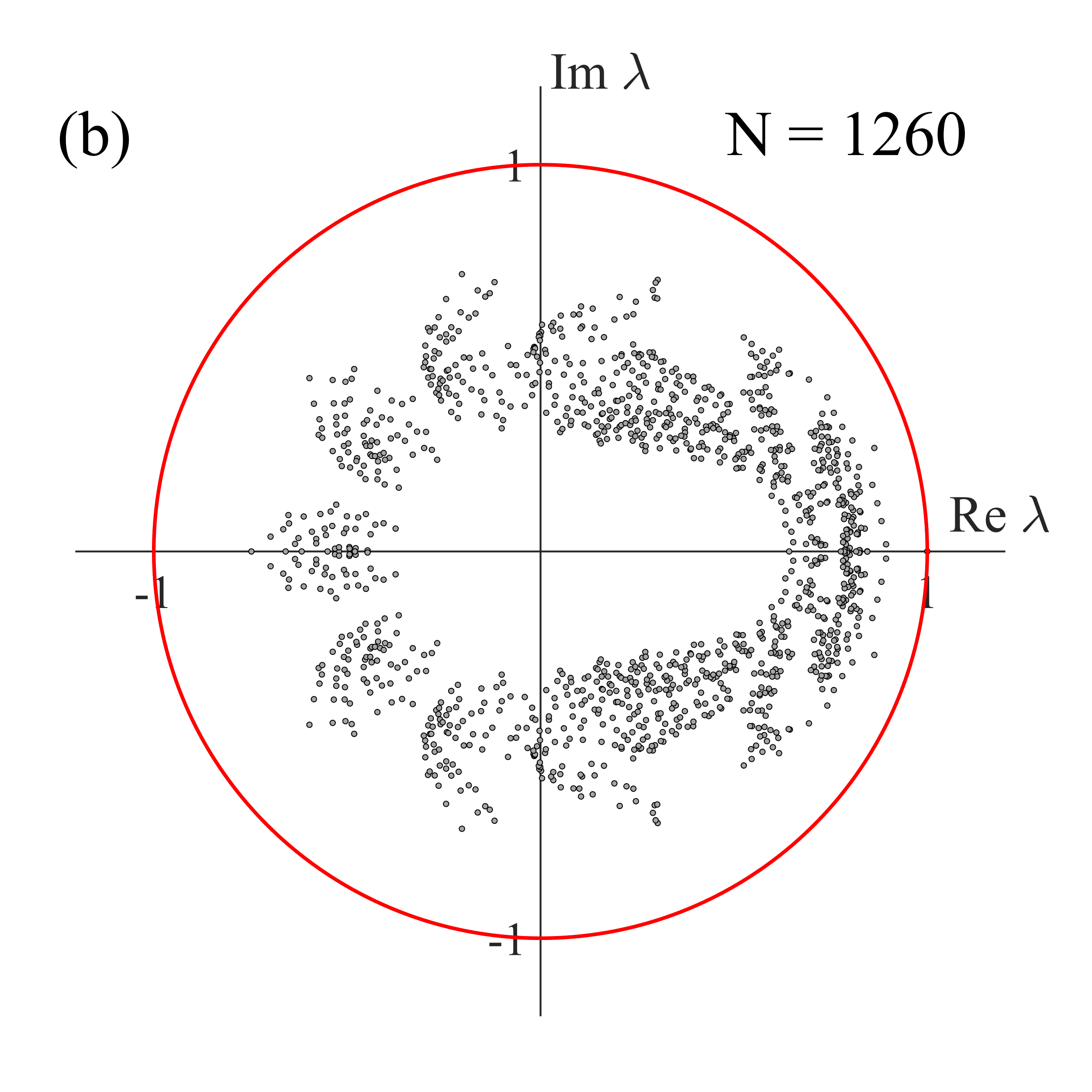}
    \end{subfigure}
    \hfill
    \begin{subfigure}[b]{0.32\textwidth}
        \includegraphics[width=\textwidth]{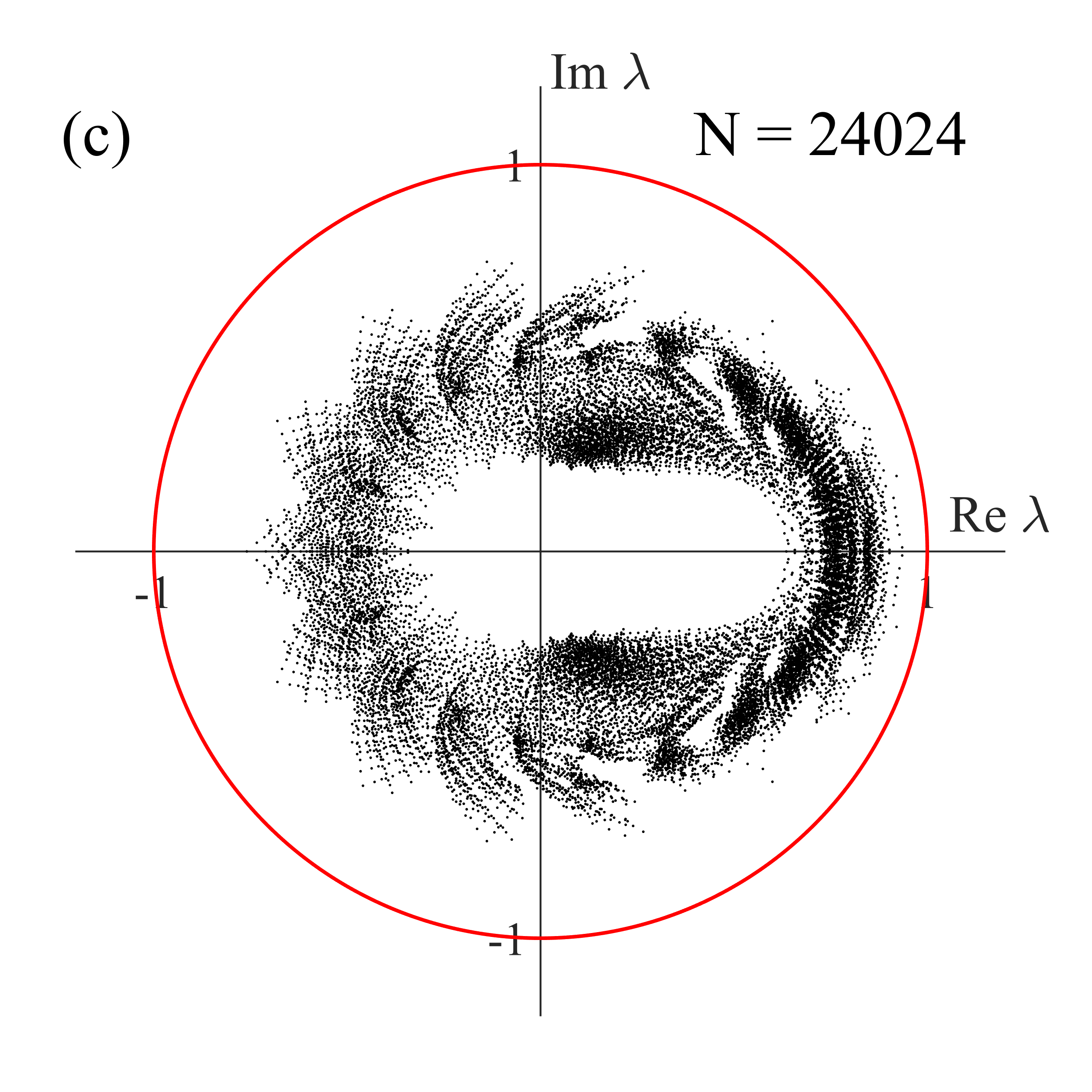}
    \end{subfigure}
    \caption{Spectra of the lifted TASEP model~\cite{Essler} for different numbers of particles $d = 3$ (b), $d = 5$ (c), and $d = 7$ (d). The  total dimension of the state space is $N = \binom{2d}{d} \cdot d$. \label{F-6}}
\end{figure}

We illustrate the above statement with the recently introduced lifted TASEP model~\cite{Essler}. In Figure~\ref{F-6}, we present the spectra of the model for different numbers $d$ of sites, corresponding to different numbers of states $N = \binom{2d}{d} \cdot d$. While we omit the details of the model formulation here, referring to the original work~\cite{Essler} for a full model description, we emphasize that the spectral supports in this case provide a clear demonstration of the absence of asymptotic localization.

Although the lifted TASEP model may appear to be an extreme case, as its eigenvalues can be expressed using the Bethe ansatz~\cite{Essler} in the form of a functional recurrence relation, the absence of asymptotic localization also extends to the standard time-discrete TASEP model~\cite{Schutz}, even in the non-integrable regime.

Although the Karpelevi\u{c} bounds are absent in the case of quantum stochastic maps, the scenario
similar in that the sampling over random maps does not resolve the entire unit disk -- even for $N=2$~\cite{BCSZ09}. Notably, while there are different methods to parameterize and sample these maps, the scenario is universal: As $N$ increases, the spectral support localizes within a disk of radius $1/N$~\cite{BCSZ09}.

\begin{figure}[t]
    \centering
    \begin{subfigure}[b]{0.32\textwidth}
        \includegraphics[width=\textwidth]{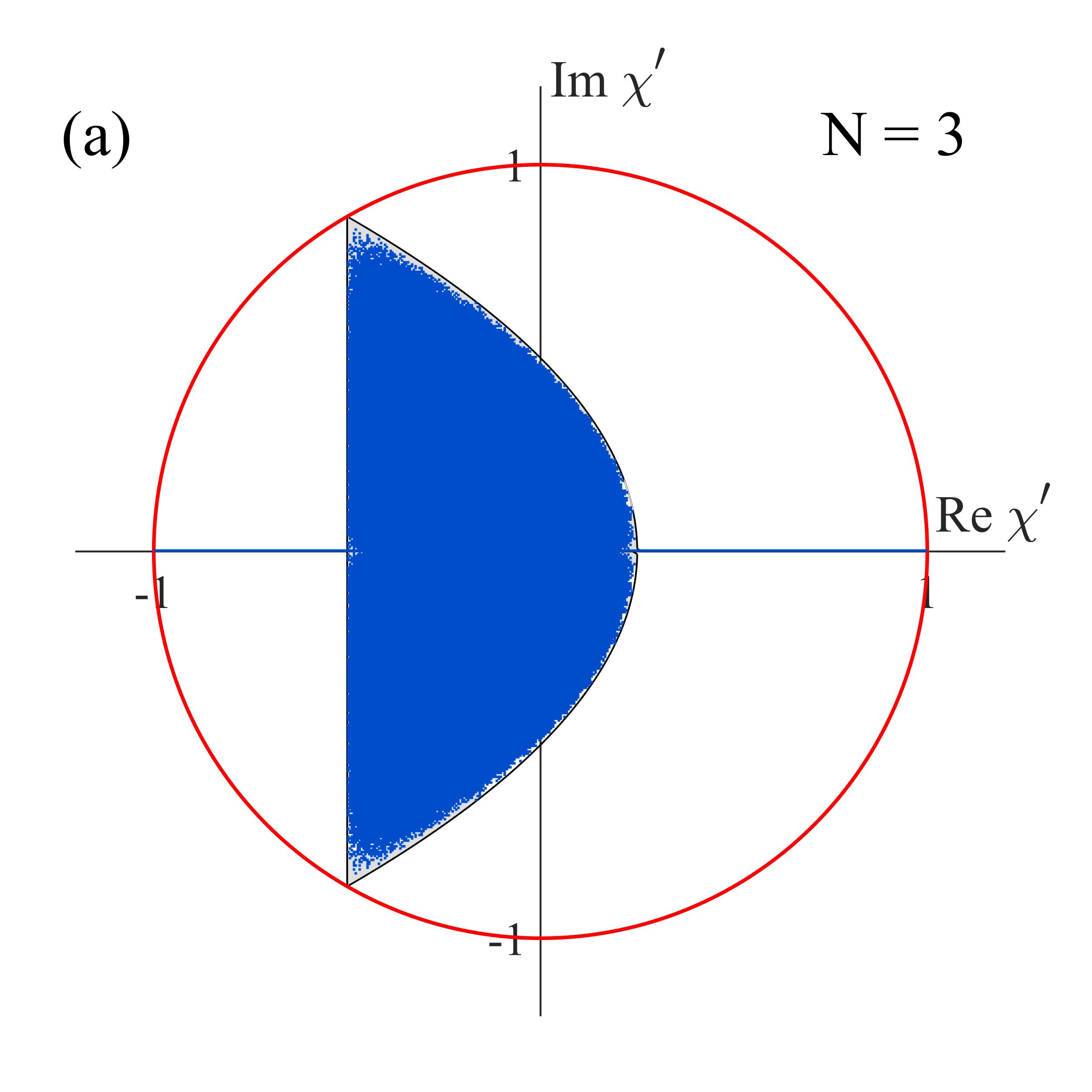}
    \end{subfigure}
    \hfill
    \begin{subfigure}[b]{0.32\textwidth}
        \includegraphics[width=\textwidth]{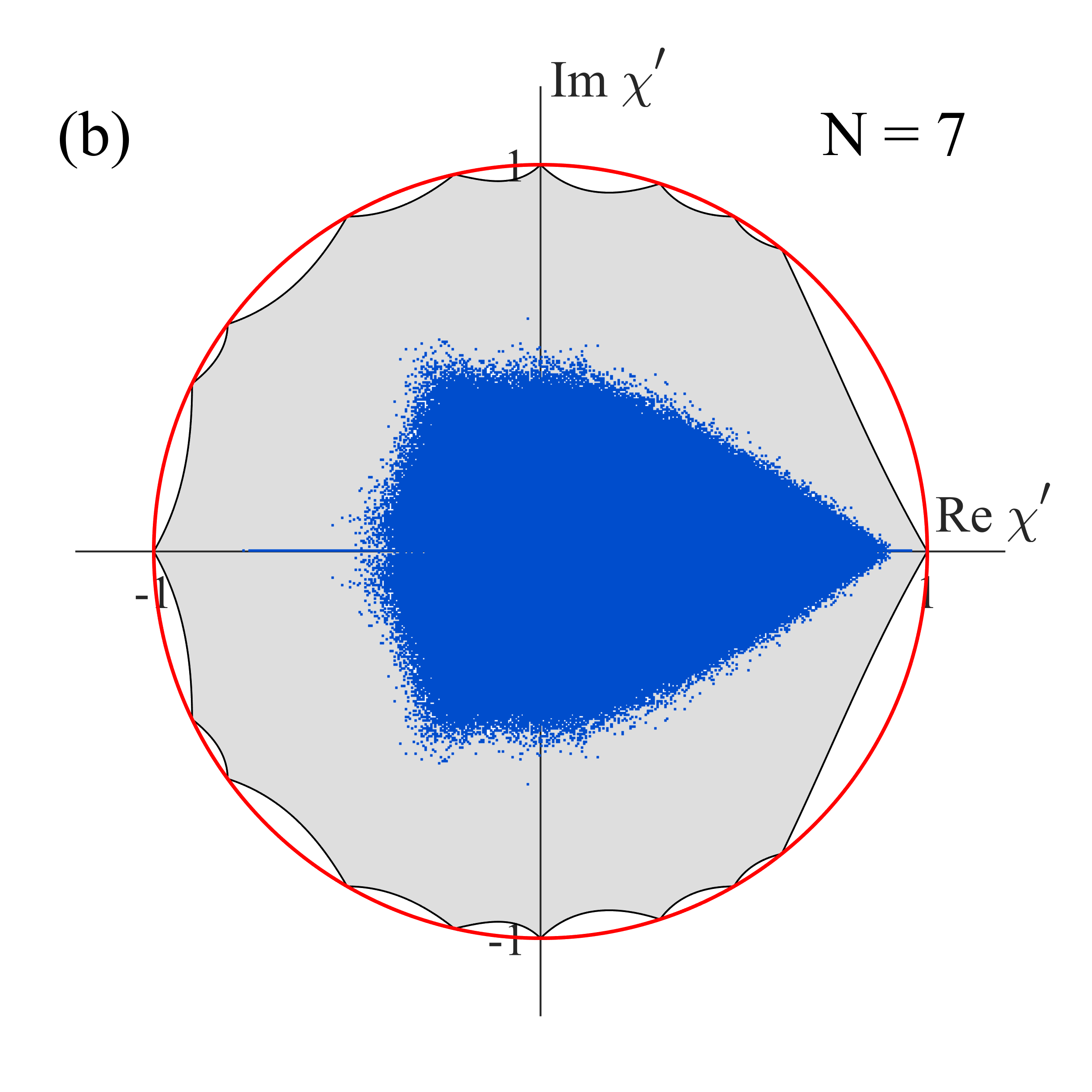}
    \end{subfigure}
    \hfill
    \begin{subfigure}[b]{0.32\textwidth}
        \includegraphics[width=\textwidth]{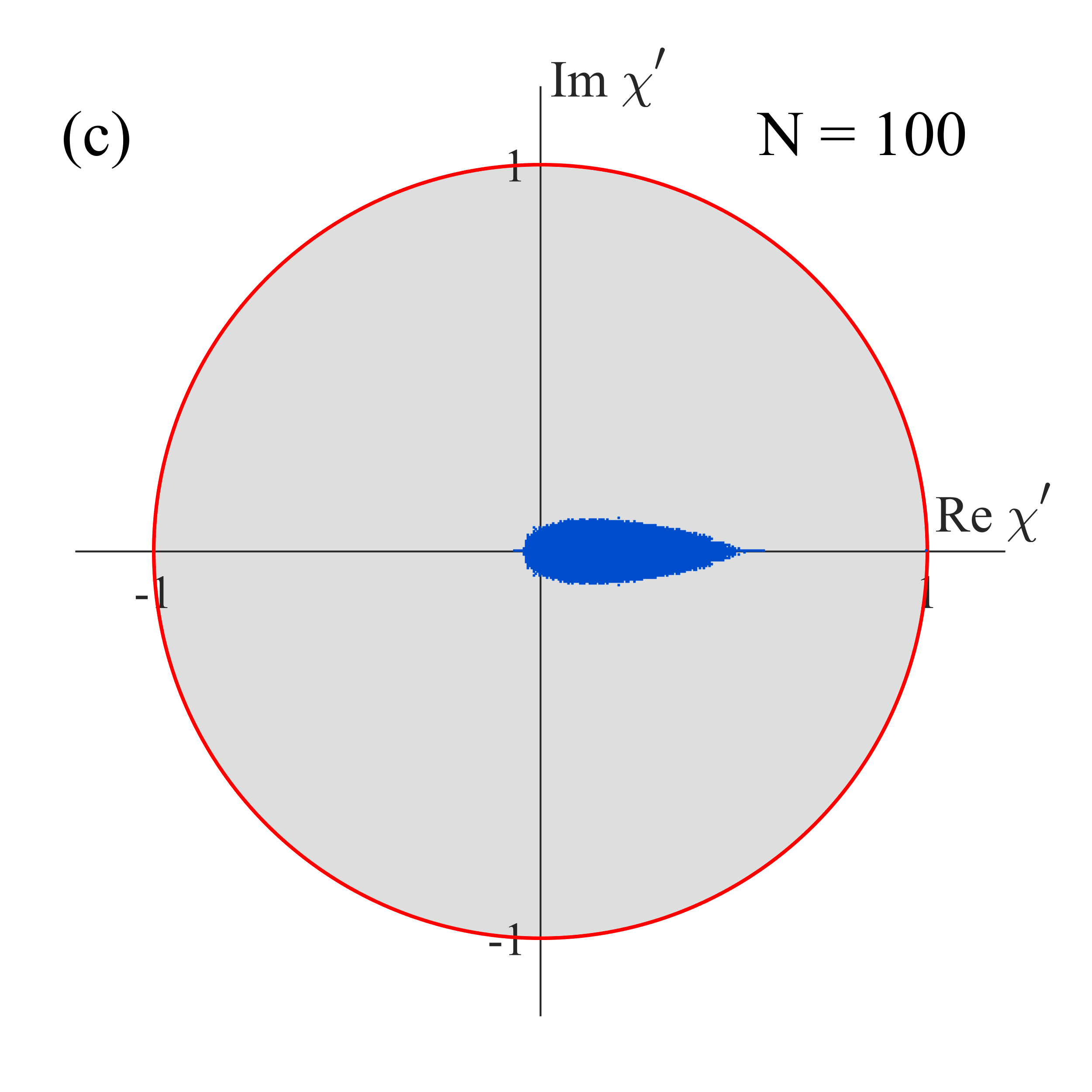}
    \end{subfigure}
    \caption{Spectral supports of rescaled Kolmogorov operators, $\tilde{\mathcal{K}}= \mathcal{K}/\nu_{\text{c}}+\oper_N$, with $\mathcal{K}$ sampled randomly, for different $N$. Results of the sampling are presented as $400 \times 400$ histograms (dark blue), where only bins containing at least one eigenvalue are shown. The number of samples is $10^8$, $10^6$, and $10^5$ for $N=3$ (a), $N=7$ (b), and $N=100$ (c), respectively. Modified Karpelevi\u{c} regions $\tilde{\Theta}_N$ are indicated in light gray.
\label{F-7}}
\end{figure}

\begin{figure}[h]
    \centering
    \begin{subfigure}[b]{0.32\textwidth}
        \includegraphics[width=\textwidth]{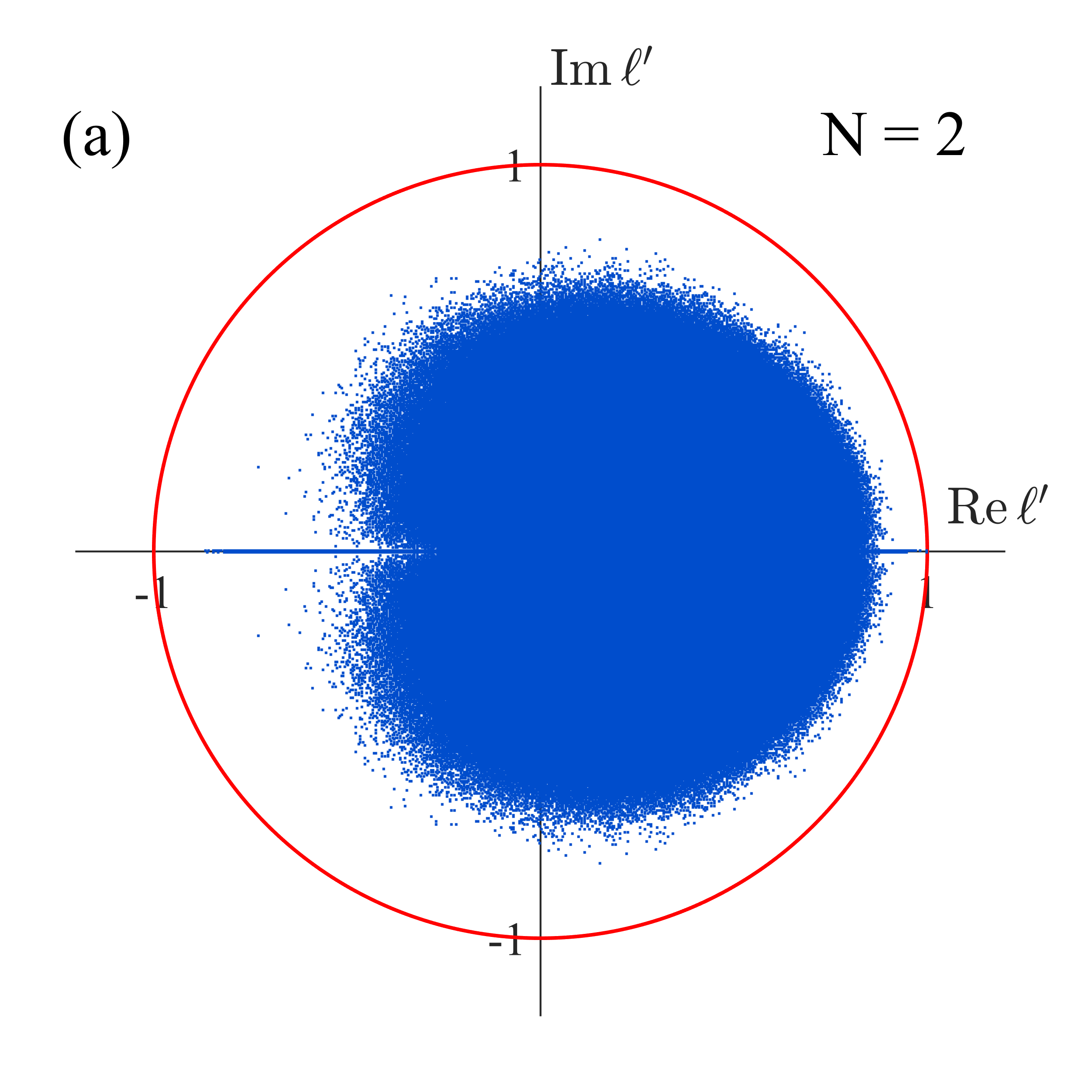}
    \end{subfigure}
    \hfill
    \begin{subfigure}[b]{0.32\textwidth}
        \includegraphics[width=\textwidth]{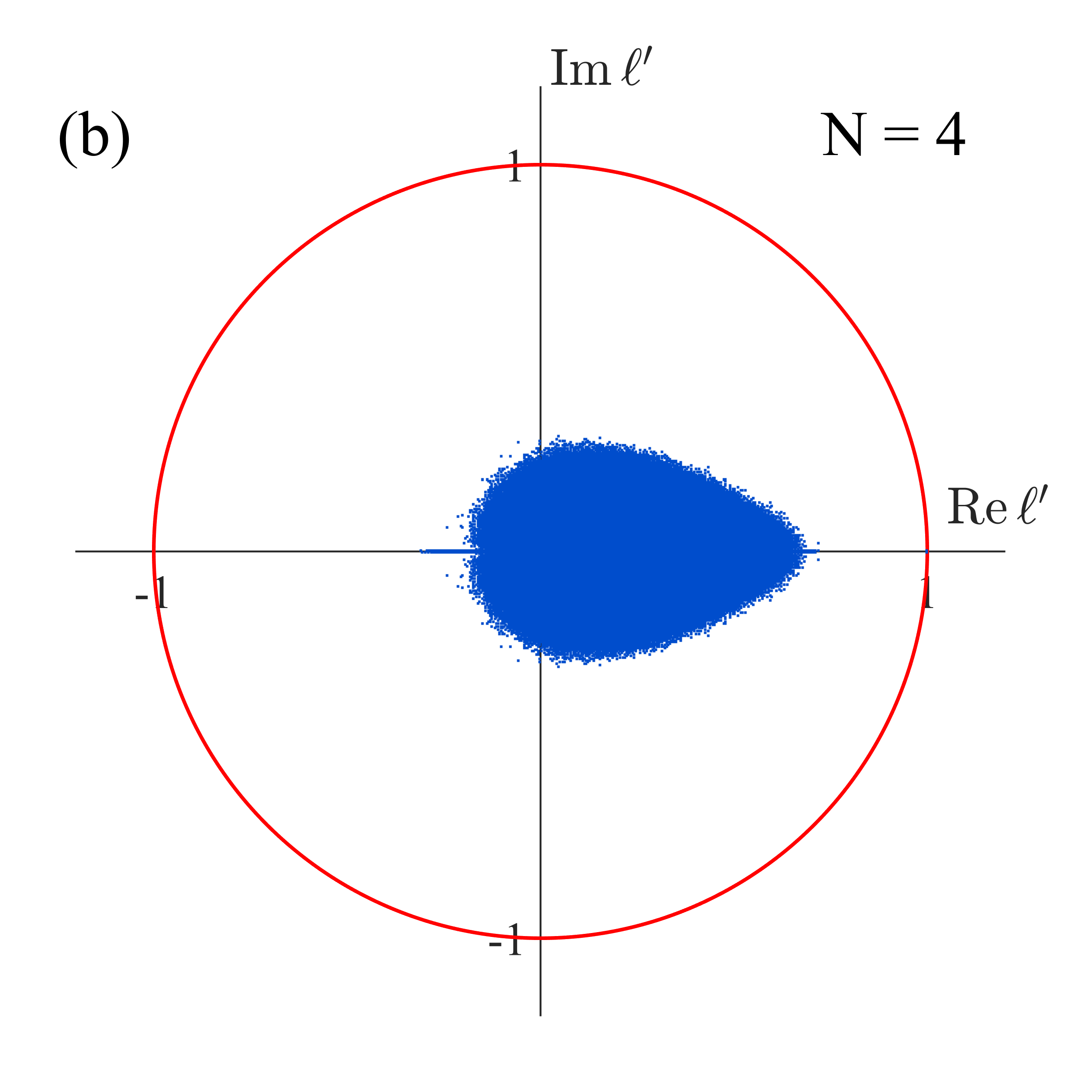}
    \end{subfigure}
    \hfill
    \begin{subfigure}[b]{0.32\textwidth}
        \includegraphics[width=\textwidth]{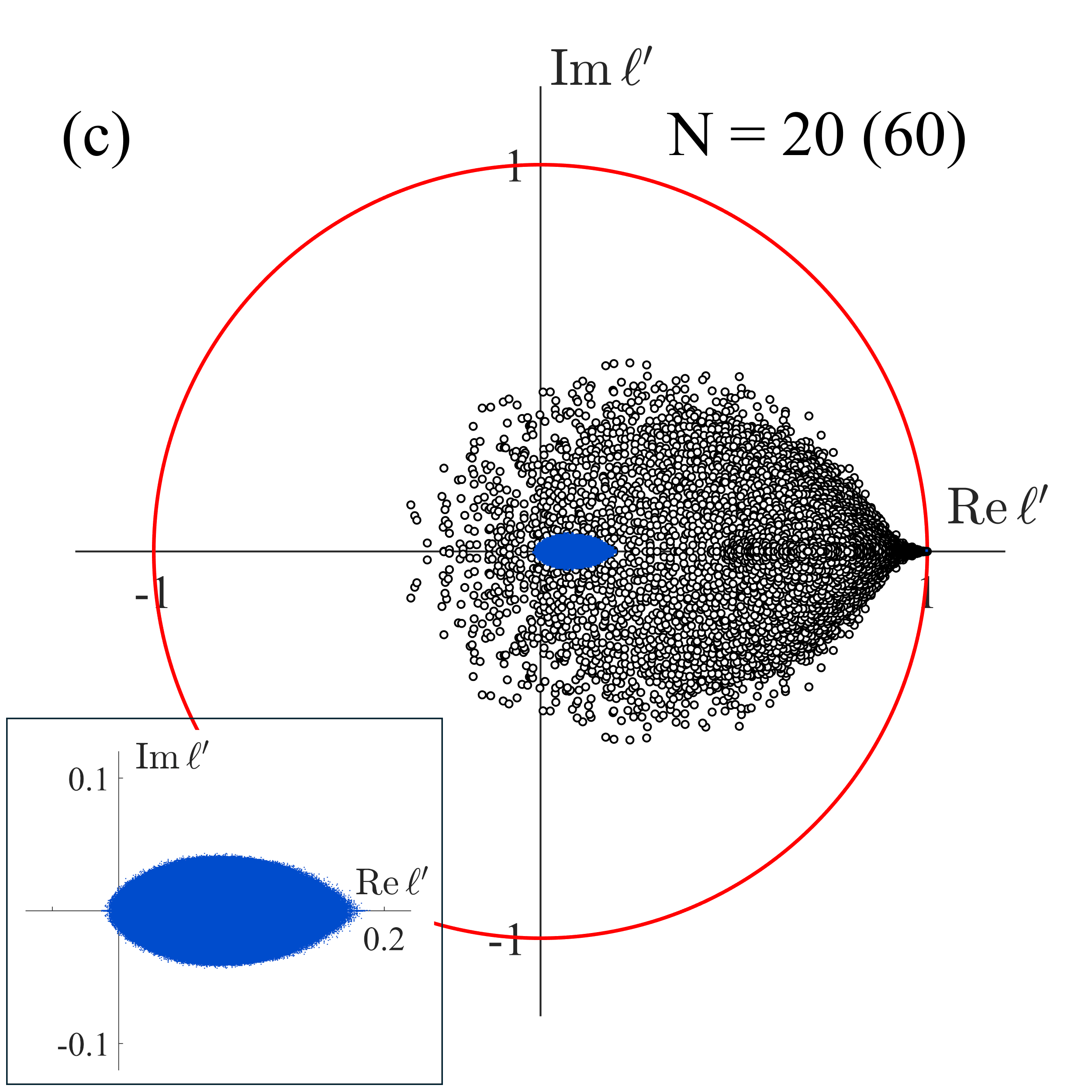}
    \end{subfigure}
    \caption{Spectral supports of rescaled Lindblad operators, $\tilde{\mathcal{L}}=  \mathcal{L}/\nu_{\text{q},2}+ {\rm id}_N$, with $\mathcal{L}$ sampled randomly, for different $N$. Results of the sampling are presented as $400 \times 400$ histograms (dark blue), where only bins containing at least one eigenvalue are shown. The number of samples is $10^8$, $10^6$, and $10^5$ for $N=2$ (a), $N=4$ (b), and $N=20$ (c), respectively. The spectrum of a rescaled Lindblad operator, $\mathcal{L}(\rho) = \Phi(\rho) - \frac{1}{2} \{ \Phi^\ddag(\oper),\rho\}$, with the map $\Phi$ obtained by performing supercoherification~\cite{Random-2} on a stochastic matrix of the lifted TASEP process~\cite{Essler} for $d=3$, $N= \binom{2d}{d} \cdot d = 60$ (see Fig.~\ref{F-6}), is shown with gray dots. 
\label{F-8}}
\end{figure}

\begin{figure}[h]
\begin{center}
\includegraphics[width=0.45\textwidth]{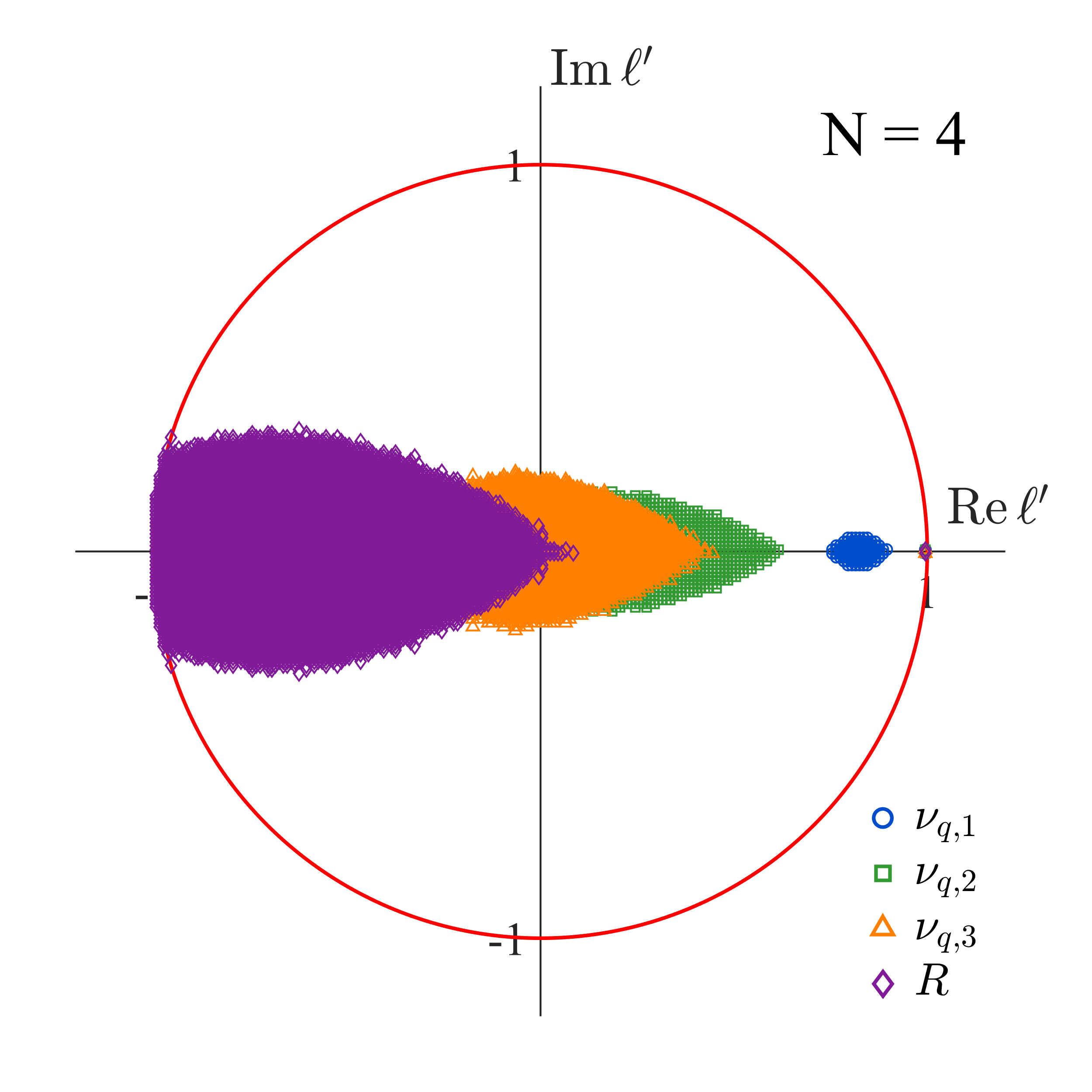}
  \caption{Spectral support of random Lindblad operators, Eq.~(\ref{L0}), for  $N=4$ and four different scaling parameters, 
  including the tight scaling parameter $R$,  Eq.~(\ref{R}).
Results of random sampling are presented as $400 \times 400$ histograms, where only bins containing at least one eigenvalue are shown. The number of samples is $10^5$. Results
    for optimal scaling parameter $\nu_{q,3}$, Eq.~(\ref{kappa}), are shown in orange.
\label{F-10}}
\end{center}
\end{figure} 

Turning now to Kolmogorov operators, we find a situation similar to that observed for stochastic matrices. Specifically, for $N=3$, the region $\tilde{\Theta}_3$ can be resolved through sampling over random Kolmogorov operators using, e.g., Eq.~(\ref{KW}); see Fig.~\ref{F-7}(a). However,   this becomes infeasible already for $N=7$; see Fig.~\ref{F-7}(b). The universality shows up 
such that, in the limit $N \gg 1$~\cite{Random-1}, the spectral support of $\sqrt{N}({\cal K} + \oper_N)$ acquires an $N$-independent spindle-like shape~\cite{timm,Random-1,Random-2}. The spectral support of the rescaled Kolmogorov operator  $\widetilde{{\cal K}} := \nu_{\text{c}}^{-1} {\cal K}$ also converges to this universal shape in the limit $N \to \infty$.  If we sample Kolmogorov operators and normalize them such that $\text{Tr} \, | \mathcal{K} | = N$~\cite{Random-2}, then in the asymptotic limit we obtain $\nu_c \to 1 + \mathcal{O}\bigl(\sqrt{\log N / N}\bigr)$. Note that $N = 100$, the value used to obtain the spectral support shown in Fig.~\ref{F-7}(b), is not large enough to reach the asymptotic distribution (in fact, it is not even sufficient to resolve the spindle).

Quite naturally, for rescaled random Lindblad operators, ${\cal L}/\nu_{\text{q},2} + {\rm id}_N$, we observe the same scenario. But in this case the sampling~\cite{Random-1} fails to resolve the unit disk already for  $N=2$, see see Fig.~\ref{F-8}(a), and the spectral density localizes much faster than in the case of Kolmogorov operators, scaling as $1/N$ (vs $1/\sqrt{N}$, as in the classical case); see Fig.~\ref{F-8}(c). 

In the asymptotic limit, the spectral support of $N({\cal L} + {\rm id}_N)$ converges to a universal, $N$-independent lemon-like shape~\cite{Random-1}. 
Similar to the case of random Kolmogorov operators, in the limit $N \to \infty$, the spectral support of  renormalized Lindblad operator, ${\cal L}/\nu_{\text{q},k} + {\rm id}_N$, converges to the universal lemon shape.

Thus, we arrive at the same conclusion we made before for  random stochastic maps: In order to escape the spell of localization, we must use (or construct) atypical Markov generators, that are, e.g., generators respecting topological constraints (for example, by connecting states arranged in chains, lattices, etc.) and/or having correlated values of transition rates. In short, the corresponding models have to originated from  natural -- physical, chemical, biological, financial, etc -- stochastic processes~\cite{anderson,lipan,sevier,grasselli,Prosen2020}.

To illustrate the statement, we again use the lifted TASEP map~\cite{Essler} with $d=3$ and, using Eq.~(\ref{KW}), construct a Kolmogorov operator from it. We then apply the procedure of \textit{supercoherification}~\cite{Random-2}, which, via the Choi-Jamiołkowski isomorphism~\cite{watrous2018,holevo,WOLF,Karol}, allows us to sample a Lindblad operator from a Kolmogorov operator.

It is noteworthy that supercoherification works in the direction opposite to that specified by Eq.~(\ref{L-K}) (the latter is referred to as "super\textit{de}coherification" in Ref.~\cite{Random-2}). Supercoherefication is a quotient operation: While, for a fixed basis, a given quantum Markov generator, when super\textit{de}coherified, gives rise to a unique classical Markov generator (for a given basis), there exist infinitely many quantum operators that can result in a given classical operator.

By performing supercoherification once, we obtain a Lindblad operator whose spectrum is shown in Fig.~\ref{F-8}(c; gray dots). Notably, even for a large dimension, $N = 60$, the spectrum does not localize and covers a substantial portion of the unit disk.

Finally, we demonstrate that the ordering of scaling parameters, $\nu_{\text{q},1} \geq \nu_{\text{q},2} \geq \nu_{\text{q},3}$, holds for random Lindbladians.  We follow the procedure from Refs.~\cite{Random-1,Random-2} and sample Lindbladians using the representation
\begin{eqnarray}  \label{L0}
\mathcal{L}(\rho) =   \sum\limits_{\alpha,\beta=1}^{N^2-1} \!\! K_{\alpha\beta} \Big[F_{\alpha}\rho F^{\dagger}_{\beta} - \frac{1}{2} \big(F^{\dagger}_{\beta} F_{\alpha}\rho + \rho F^{\dagger}_{\beta} F_{\alpha}\big)\Big], \qquad
\end{eqnarray}
where  $\{F_\alpha\}$ is the set of the generalized  Gell-Mann matrices~\cite{ALICKI}.

With this representation, we can sample random Lindbladians by drawing Kossakowski matrices from an ensemble of complex Wishart matrices with trace $N$,
\begin{equation}
K = N \, G G^\dagger / \mathrm{Tr}(G G^\dagger),
\label{eq:RandomKoss}
\end{equation}
where $G$ is a complex Ginibre matrix with i.i.d. complex Gaussian entries.

For such an ensemble of Lindbladians, the value of $\nu_{\text{q},1}$, Eq.~(\ref{norm1}), is sample-independent and equals $N$, due to the normalization of the Kossakowski matrix. The quantity $\nu_{\text{q},2}$ is computed according to its definition, Eq.~(\ref{nunu}). For each sampled Lindbladian, we estimate $\nu_{\text{q},3}$, Eq.~(\ref{kappa}), by evaluating $|\langle \psi | \mathcal{L}(|\psi\rangle\langle\psi|) | \psi \rangle|$ over $10^4$ Haar-random pure states.
The maximal value is then taken as $\nu_{\text{q},3}$. We also 
use tight scaling $R$,  Eq.~(\ref{R}), to map the results of the $\nu$-scalings onto it.


The results of spectral sampling for $N = 4$ (with $10^5$ samples), rescaled using different scaling constants, $\nu_{\text{q},k}$, $k \in \{1,2,3\}$, and $R$, are presented in Figure~\ref{F-10}. The plot illustrates the hierarchy in a clear-cut way. Moreover, the hierarchy is not violated in a single instance throughout the entire ensemble of $10^5$ samples.

\section{Conclusions}  
 \label{Conclusions}

We analyzed the spectra of the classical, ${\cal K}$, and quantum, $\mathcal{L}$, Markov generators. 
We introduce a rescaling procedure, for both limits, such that the spectra of rescaled generators are confined to the unit disk $\mathbb{D}(-1,1)$. 

\begin{figure}[t]
\begin{center}
\includegraphics[width=0.99\textwidth]{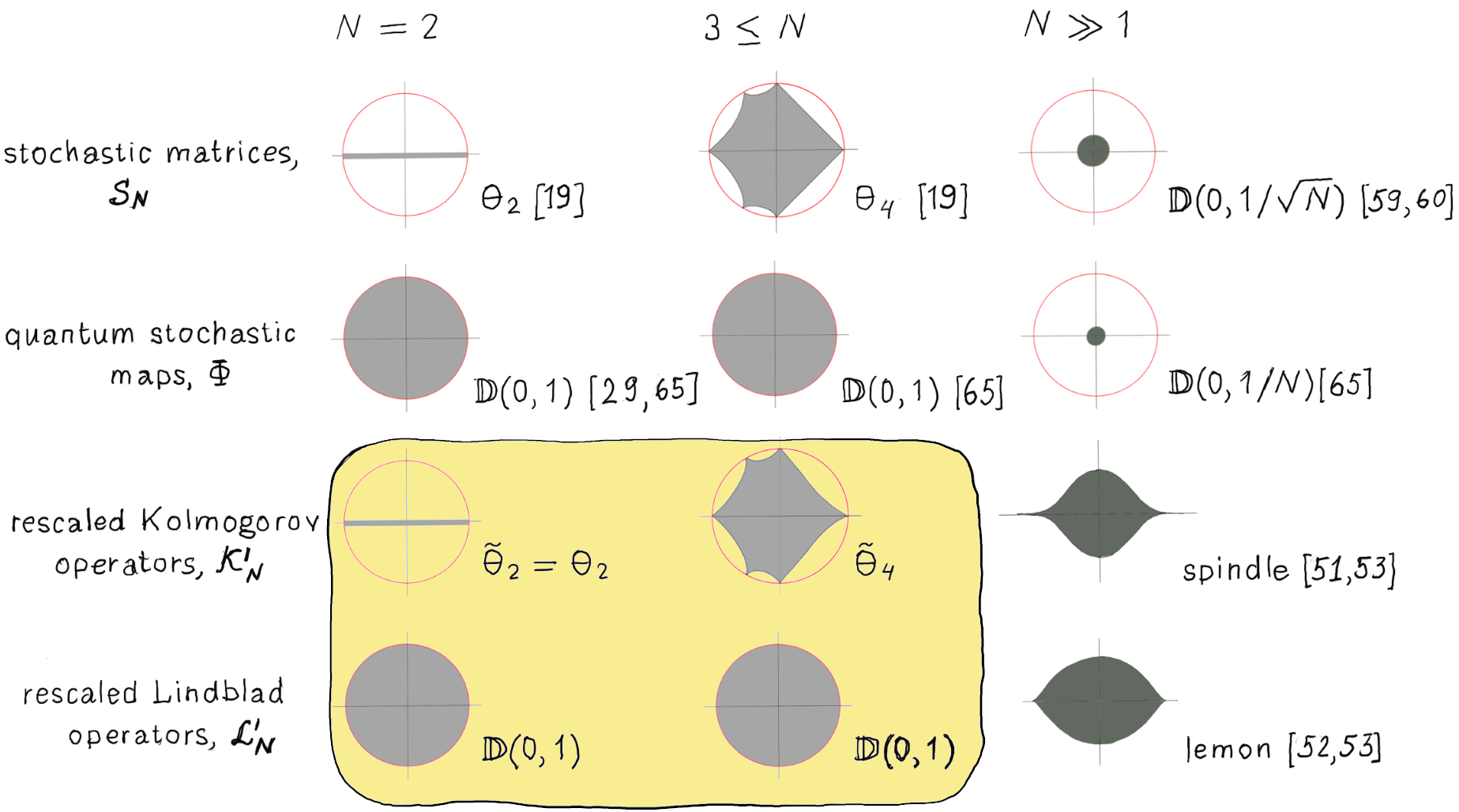}
\caption{Delineation of the spectra for stochastic matrices $S_N$, quantum stochastic maps $\Phi_N$, and rescaled classical, $K'$, and quantum, ${\cal L}'$, Markov generators. Each spectrum contains the leading eigenvalue $\lambda_0 = 1$, and the remaining bulk. We address the latter (marked gray) in this sketch.
In the first two columns (from the left), we address the case of spectral bounds for \textit{all} possible operators and consider the classical [quantum] generators scaled as $K' = K / \nu_{\text{c}} + \oper_N$ [${\cal L}' = \mathcal{L} / \nu_{\text{q}} + {\rm id}_N$, $\nu_{\text{q}} \in \{\nu_{\text{q},1}, \nu_{\text{q},2}, \nu_{\text{q},3} \}$].  
In the last column, we focus on the case of \textit{random (typical)} operators, and consider scaling of classical [quantum] generators as $K' = \sqrt{N}({\cal K} + \oper_N)$ [${\cal L}' = N({\cal L} + {\rm id}_N)$], with ${\cal K}$ and ${\cal L}$ sampled randomly (see Section~\ref{Sec_typical}).
\label{F-11}}
\end{center}
\end{figure}

Moreover, we demonstrated that, for a given dimension $N$, the spectra of rescaled Kolmogorov operators, ${\cal K}'= {\cal K} /\nu_{\text{c}}+\mathbb{I}_N$, are confined to the modified Karpelevi\u{c} regions, ${\tilde{\Theta}}_N \subset \Theta_N \subset \mathbb{D}(0,1)$,  where the original Karpelevi\u{c} region $\Theta_N$ bounds the spectra of stochastic $N \times N$ matrices (see Fig.~\ref{F-0}).  

It should be emphasized that this result is universal and applies to all classical Markov generators with an arbitrary finite number of states $N$. Moreover, the rescaling parameter  
$\nu_{\text{c}}$ is optimal in the sense that it is  
the smallest scaling parameter that guarantees the confinement of the spectra to the unit disk.

In the quantum setting, we studied the spectra of Markovian generators 
$\mathcal{L}$, which describe continuous--time quantum evolution governed 
by the Gorini--Kossakowski--Lindblad--Sudarshan equation~(\ref{GKLS}). 
We proposed three different scaling parameters $\nu_{\text{q},k}$, $k=1,2,3$, and conjectured (Conjectures \ref{CON-0}--\ref{CON-3})  that the spectrum of a rescaled generator 
$\mathcal{L}' = \mathcal{L}/\nu_{\text{q}} + {\rm id}_N$, 
with the scaling parameter $\nu_{\text{q}} \in \{\nu_{\text{q},1},\nu_{\text{q},2},\nu_{\text{q},3}\} $, 
is confined to the unit disk. It is also conjectured that $\nu_{\text{q},3}$ provides an optimal scaling,  $\frac{1}{\nu_{\text{q},3}}\mathcal{L}$, and cannot be bounded any further. These conjectures are supported both by numerical simulations and by examining several Lindblad generators commonly studied in the literature~\cite{ALICKI,Breuer,rivas,Bassano}.

The sketch in Fig.~\ref{F-11} summarizes these findings (highlighted in frame) by placing them within the existing context of knowledge about the spectra of stochastic maps and generators.


In Section~\ref{Beyond}, we went beyond the limit of completely positive semigroups, discussed generators of PTP evolution, and proposed 
an intrinsic definition of the rescaling parameter $\nu_{\text{q},3}$ (see Conjecture~\ref{CON-3}). Interestingly, an analysis of simple qubit examples suggests that $\nu_{\text{q},3}$ indeed provides the minimal scaling value.

The analysis of spectral properties of Markov generators is closely related to relaxation rates, 
which are important characteristics of both classical and quantum 
processes. The relaxation rates are defined as
$\Gamma_k := - \mathrm{Re}\,\xi_k$, where $\xi_k$ are the 
eigenvalues of the corresponding Markov generator. 
In a recent work \cite{Rates}, it was shown 
that the relaxation rates of any Lindblad operator satisfy
\begin{equation}  \label{GG}
    \Gamma_k \;\le\; \frac{\Gamma}{N},
\end{equation}
where 
$\Gamma = \Gamma_1 + \cdots + \Gamma_{N^2-1}$. 
This inequality thus imposes a universal constraint on the real part 
of the spectrum. In particular, the real part of the spectrum of 
the rescaled generator $\tfrac{1}{\Gamma}\,\mathcal{L}$ 
is bounded below by $-\tfrac{1}{N}$. However, this constraint does not address the imaginary part of the spectrum and the results presented here partially fill this gap.

It is noteworthy that, since the rescalings we introduced are linear transformations, they do
not affect eigenvalue correlation measures such as the complex spacing ratio~\cite{Prosen2020}, which was recently introduced as an indicator of chaos and integrability of Markov generators. Therefore, the corresponding classification of  Kolmogorov and Lindblad operators into 
``integrable'' and ``chaotic'' remains unaffected by scalings.

As a next step, it would be interesting to generalize the analysis of the spectral bounds of Markov generators to the infinite-dimensional setting. Another important direction is the spectra of classical and quantum superchannels and the corresponding Markov super-generators. Quantum superchannels \cite{Chiribella} are a fundamental concept in quantum information theory, generalizing the notion of quantum channels to higher-order transformations; see. recent reviews~\cite{Gour,Modi}). While quantum channels represent physical processes that map quantum states to quantum states, superchannels describe transformations that act on quantum channels themselves. Similarly, classical superchannels generalize the notion of classical channels (stochastic matrices) to higher-order transformations. That is, while stochastic matrices represent physical processes that map classical states (probability vectors) to classical states,  a classical superchannel transforms stochastic matrices to other stochastic matrices. The corresponding Markov super-generators were recently studied in Ref.~\cite{Caro}. It would be highly interesting to establish spectral bounds for these higher-order operators, both in the classical and quantum settings.


\section*{Acknowledgments}
	
D. C. and W. T. were
supported by the Polish National Science Center
under Projects No. 2018/30/A/ST2/00837
and 2021/03/Y/ST2/00193, respectively.
The latter project, carried out within the QuantERA II Program project DQUANT, has also received funding from the European Union’s Horizon 2020 Research and Innovation Programme under Grant Agreement No. 101017733. K. {\.Z}. acknowledges funding from the European Union under the ERC Advanced Grant TAtypic (project number 101142236). S.D. acknowledges support from the Research Council of Norway through the project “IKTPLUSS – IKT og digital innovasjon” (No. 333979).


\appendix


\section{$N$-level Lindbladians}   \label{APP}

\counterwithin{equation}{section}
\setcounter{equation}{0}

In  this section we analyze the spectra of Lindblad operators
rescaled by the  quantum scaling parameter 
 $\nu_{\text{q}}:=\nu_{\text{q},2}$ given in Eq.~(\ref{nunu}).
Consider a map acting on an $N$ dimensional system,
sometimes called a {\sl qunit},
\begin{equation}
\Phi(X) = \sum_{i,j=1}^N W_{ij} |i\>\<j| X |j\>\<i| ,
\end{equation}
with $W_{ij}\geq 0$ and $W_{ii}=0$.  We have

\begin{equation}
    \| \Phi^\ddag\|_\infty = \| \Phi^\ddag(\oper)\|_\infty = \max_j w_j ,
\end{equation}
where $w_j = \sum_i W_{ij}$. Now, $D$ is a diagonal matrix with $D_{ii} = \nu_{\text{q}} - w_i$.  The spectrum of $\widetilde{\Phi}$ consists of $N$ eigenvalues of $\widetilde{W}_{ij}$ defined in Eq.~(\ref{tilde-W}) and
\begin{equation}
    \widetilde{\lambda}_{ij} = \nu_{\text{q}} - \frac{w_i + w_j}{2} , \ \ i \neq j .
\end{equation}
Eigenvalues of $\widetilde{W}_{ij}$ belongs to the disk of radius $\nu_{\text{c}}\leq \nu_{\text{q}}$ and $\widetilde{\lambda}_{ij} \in [-\nu_{\text{q}},\nu_{\text{q}}]$ and hence indeed the spectrum of $\widetilde{\Phi}$ belongs to the disk of radius $\nu_{\text{q}}$. 
    
Consider now the generator giving rise to the pure qunit decoherence, defined by

\begin{equation}
\Phi(X) = \sum_{i,j=1}^N A_{ij} |i\>\<i| X |j\>\<j| ,
\end{equation}
where the matrix $A_{ij}$ is positive definite. Note, that $\Phi(X) = A \circ X$ is just a Hadamard product of $A$ and $X$. One finds $\Phi^\ddag(\oper) = \sum_i A_{ii} |i\>\<i|$, and hence

\begin{equation}
   \nu_{\text{q}}:= \| \Phi^\ddag\|_\infty = \| \Phi^\ddag(\oper)\|_\infty = \max_i A_{ii} .
\end{equation}
Finally, the matrix $D$ is diagonal with $D_{ii} = \nu_{\text{q}} - A_{ii}$. One easily finds for the spectrum of $\widetilde{\Phi}$


\begin{equation}
    \widetilde{\lambda}_{ij} = A_{ij} + \nu_{\text{q}} - \frac{1}{2}( A_{ii}+A_{jj}) .
\end{equation}
To show that $| \widetilde{\lambda}_{ij} |\leq \nu_{\text{q}}$, let first us observe that 
 $$   A_{ii} = x_i \nu_{\text{q}} \ , \ \ \ A_{jj} = x_j \nu_{\text{q}} ,$$
with $x_i,x_j \in[0,1]$. Positivity of  $[A_{ij}]$ implies $A_{ii} A_{jj} \geq |A_{ij}|^2$ and hence due to $\frac{1}{2}( A_{ii}+A_{jj}) \geq \sqrt{A_{ii} A_{jj}} $ we have

$$   A_{ij} = \nu_{\text{q}} \, \frac{r_{ij}}{2}( x_{i}+x_{j}) e^{i \phi_{ij}} ,   $$
with $r_{ij} \in [0,1]$. Finally, denoting $x_{ij} := \frac 12(x_i+x_j)$ one has

\begin{equation}
    | \widetilde{\lambda}_{ij} |^2 = \nu_{\text{q}}^2 \Big| 1 - x_{ij}  + x_{ij}r_{ij}e^{i\phi_{ij}}\Big|^2 \leq \nu_{\text{q}}^2 ( 1 - x_{ij}  + x_{ij}r_{ij})^2 \leq \nu_{\text{q}}^2 .
\end{equation}

\begin{figure}[t]
\begin{center}
\includegraphics[width=0.4\textwidth]{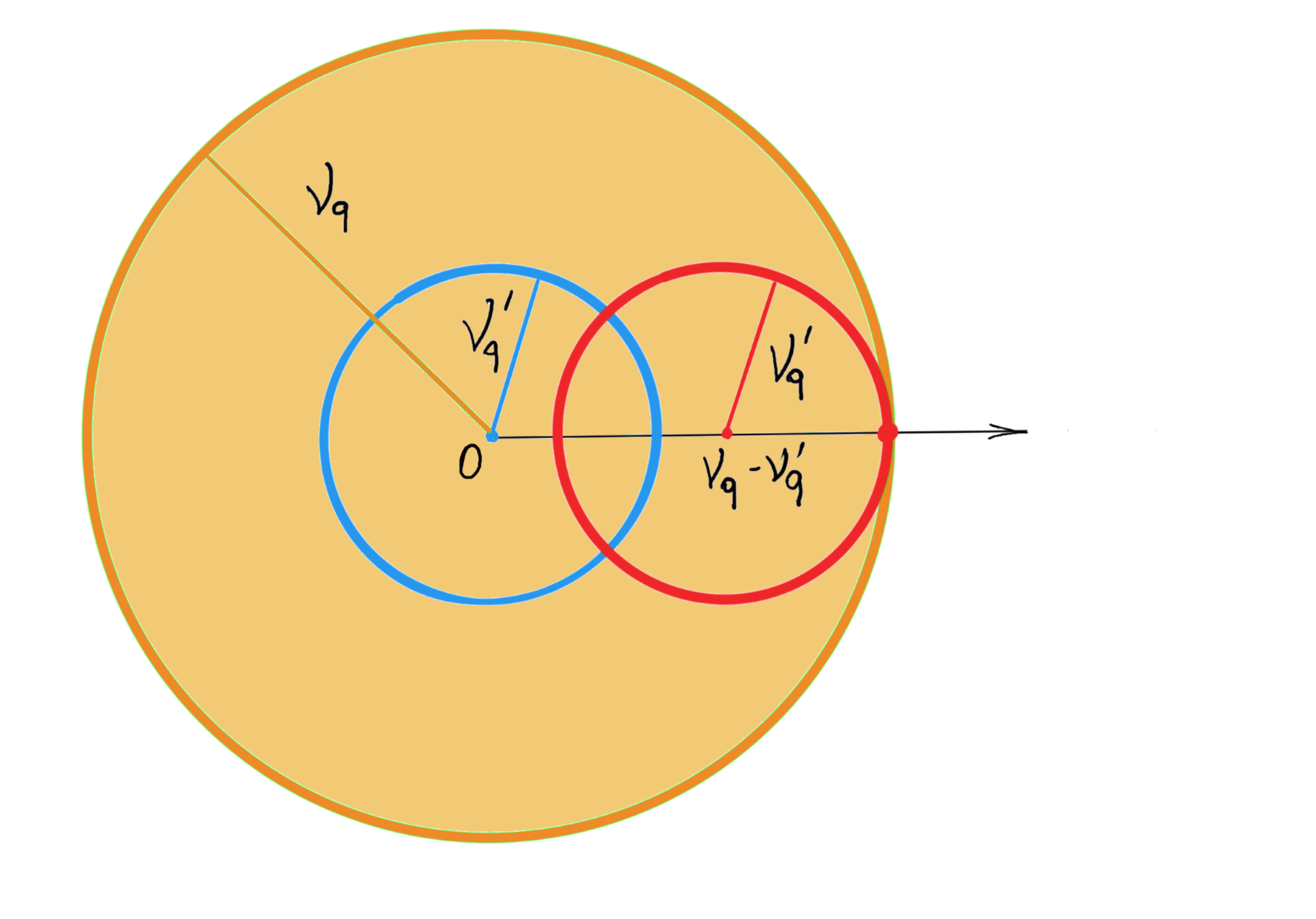}
\caption{Relation between  scaling parameters 
$\nu_{\text{q}}$ and $\nu'_{\text{q}}$.  
\label{F-I}}
\end{center}
\end{figure}

 Consider now one of the most commonly considered in the literature types of Lindblad operators, $\mathcal{L}=\mathcal{L}_1 + \mathcal{L}_2$, where

\begin{equation}
    \mathcal{L}_1(\rho) = \sum_{i\neq j} W_{ij} |i\>\<j|\rho|j\>\<i| - \frac 12 \sum_j w_j (|j\>\<j|\rho + \rho|j\>\<j|) , 
\end{equation}
with $w_j =\sum_i W_{ij}$, and

\begin{equation}
    \mathcal{L}_2(\rho) = \sum_{i,j} A_{ij} |i\>\<i|\rho|j\>\<j| - \frac 12 \sum_i A_{ii}  (|i\>\<i|\rho + \rho|i\>\<i|) .
\end{equation}
This is the most general Lindbladian covariant w.r.t. diagonal unitary matrices, i.e.

\begin{equation}
    U \mathcal{L}(\rho) U^\dagger = \mathcal{L}(U\rho U^\dagger) ,
\end{equation}
for all $U = \sum_k \e^{i \phi_k} |k\>\<k|$. In particular, the celebrated Davies generator derived in the weak coupling limit belongs to this class \cite{Davies-1,ALICKI,Breuer,rivas,Bassano,PR2022}. 
Note that $[\mathcal{L}_1,\mathcal{L}_2]=0$ and hence the spectral properties of $\mathcal{L}_1+\mathcal{L}_2$ are easy to analyze. One finds

\begin{equation}
    \|\Phi^\ddag\|_\infty = \|\Phi^\ddag(\oper)\|_\infty  = \max_i (w_i + A_{ii}) \leq \nu'_{\text{q}} + \nu''_{\text{q}}, 
\end{equation}
where $\nu'_{\text{q}} = \|\Phi_1^\ddag\|_\infty = \max_i w_i$, and $\nu''_{\text{q}} = \|\Phi_2^\ddag\|_\infty = \max_i A_{ii}$. One easily finds for the corresponding map $\widetilde{\Phi}$

\begin{equation}
    \widetilde{\Phi}(X) = \sum_{i\neq j} W_{ij} |i\>\<j|X|j\>\<i| + \sum_{i,j} A_{ij} |i\>\<i|\rho|j\>\<j| + \nu_{\text{q}}\, X - \frac 12 \sum_i (w_i+A_{ii}) (|i\>\<i|X + X|i\>\<i|) .
\end{equation}
Again, the spectrum consists of $N$ eigenvalues of $\widetilde{W}_{ij}$ defined in (\ref{tilde-W}) located at the disk of radius $\nu'_{\text{q}}$ and shifted by $\nu_{\text{q}}-\nu'_{\text{q}}$, i.e., they are located at the disk of radius $\nu'_{\text{q}}$ located at $(\nu_{\text{q}}-\nu'_{\text{q}},0)$ which defines a subset of a disk of radius $\nu_{\text{q}}$ located at the center, see Fig. \ref{F-I}. The remaining eigenvalues read

\begin{equation}
    \widetilde{\Phi}(|i\>\<j|) = \widetilde{\lambda}_{ij} |i\>\<j| ,\ \ \ \ (i\neq j)
\end{equation}
with

\begin{equation}
     \widetilde{\lambda}_{ij} = A_{ij} + \nu_{\text{q}} - \frac 12 (w_i+w_j+A_{ii}+A_{jj}) . 
\end{equation}
Similar analysis  shows that $|\widetilde{\lambda}_{ij}|\leq \nu_{\text{q}}$.

\end{document}